\documentclass[useAMS,usenatbib,usegraphicx]{./mn2e}
\usepackage{natbib}
\usepackage{lscape}

\newcommand{\Teff}{\mbox{$T_{\rm eff}$}~}
\newcommand{\teff}{\mbox{$T_{\rm eff}$}~}

\newcommand{\Msun}{\mbox{$M_{\odot}$}}

\newcommand{\as}{\mbox{$^{\prime\prime}$~}}

\newcommand{\beq}{\begin{equation}}
\newcommand{\eeq}{\end{equation}}
\newcommand{\beqa}{\begin{eqnarray}}
\newcommand{\eeqa}{\end{eqnarray}}

\newcommand{\diff}{\mbox{${\rm d}$}}

\newcommand{\ebv}{\mbox{$E_{B\!-\!V}$}}

\title[Hot White Dwarfs: UV catalogs]{Catalogs of  Hot White Dwarfs in the Milky Way from GALEX's  Ultraviolet Sky Surveys.
Constraining Stellar Evolution.}

\author[L.Bianchi et al]{{Luciana Bianchi$^{1}$\thanks{E-mail: bianchi@pha.jhu.edu}, 
Boryana Efremova$^{1}$,  
James Herald$^{1}$, 
L\'eo Girardi$^{2}$,}  
\newauthor{Alexandre Zabot$^{3}$, 
Paola Marigo$^{4}$ and  
 Christopher Martin$^{5}$}\\
$^{1}${Department of Physics and Astronomy, Johns Hopkins University, 3400 N.Charles St.,  Baltimore, MD 21218, USA; bianchi@pha.jhu.edu}\\
$^{2}${Astronomical Observatory of Padova, INAF, Vicolo dell'Osservatorio 5, I-35122 Padova,  Italy}\\
$^{3}${Universidade Federal da Fronteira Sul, Campus de Laranjeiras do Sul, Brazil}\\
$^{4}${Dept. of Astronomy, University of Padua, Vicolo dell'Osservatorio 3, I-35122, Padova. Italy}\\
$^{5}${CA Institute of Technology, Pasadena, USA} }
\begin{document}

\date{Accepted  ----  Received \today} 

\maketitle

\label{firstpage}

\begin{abstract}

We present  comprehensive catalogs of hot star candidates in the Milky Way, selected from GALEX far-UV (FUV, 1344-1786\AA) 
and near-UV (NUV, 1771-2831\AA)  imaging.  The FUV and NUV photometry 
allows us to extract the hottest stellar objects, 
in particular hot white dwarfs (WD),  which are elusive at other wavelengths because of their
 high temperatures  and faint optical luminosities.
We generated  catalogs of UV sources from two GALEX's surveys: AIS (All-Sky Imaging Survey, depth 
ABmag$\sim$19.9/20.8 in FUV/NUV) and MIS
(Medium-depth Imaging Survey, depth $\sim$22.6/22.7mag). The two catalogs 
(from GALEX fifth data release) contain  65.3/12.6~million  (AIS/MIS) 
unique UV sources with error$_{NUV}$$\leq$0.5mag, over 21,435/1,579 square degrees.
We also constructed subcatalogs of the UV sources with matched optical photometry from SDSS (seventh data release): these contain 
0.6/0.9million (AIS/MIS) sources with errors $\leq$0.3mag in both FUV and NUV,  excluding sources with multiple optical counterparts,
 over  an area of 7,325/1,103 square degrees. All catalogs are available online. 
   We then selected  28,319~(AIS)~/~9,028~(MIS)  matched sources with FUV-NUV$<$-0.13; 
 this color cut corresponds to stellar \teff  hotter than $\sim$18,000~K
(the exact value varying with gravity).
An additional color cut of NUV-{\it r}$>$0.1  isolates binaries with largely differing \Teff's, and 
some intruding QSOs (more numerous at faint magnitudes). 
Available spectroscopy for a subsample
indicates that  hot-star candidates with NUV-{\it r}$<$0.1 (mostly ``single'' hot stars) have 
 negligible contamination by non-stellar objects. 
We discuss the distribution of sources in the catalogs, and the effects of error and color cuts on the samples.
The density of hot-star candidates increases from high to low Galactic latitudes, 
 but drops on the MW plane due to dust extinction. 
Our hot-star counts 
at all latitudes are better matched by Milky Way models computed with an initial-final mass relation (IFMR) that favours lower final masses.
The model analysis indicates that the brightest  sample is likely 
composed of WDs located in the thin disk, at typical distances between 0.15-1kpc, while the fainter sample
comprises also a fraction of thick disk and halo stars. Proper motion
distributions, available only for the bright sample (NUV$<$18~mag), are consistent with the kinematics of a thin-disk population. 
\end{abstract}


\section{Introduction}
\label{s_intro}

The vast majority of stars (initial mass 
$\la$8\Msun)~ end their lives as white dwarfs (WD),
after passing through the asymptotic giant branch (AGB) and planetary nebula (PN) phases,
in which they shed much of their mass.
The ejected material enriches
the interstellar medium (ISM) with
newly synthesized nuclear products (mainly He, C, N, and possibly O),
to different extents, depending on the initial stellar mass and
exact evolutionary path (e.g. Marigo 2001, Karakas 2010).
Intermediate mass stars are 
the main providers of carbon and nitrogen, 
whereas low-mass stars are the most relevant component for the mass budget of stellar remnants in galaxies. 
Most of the stellar mass is shed  in the AGB and PN phases, but the evolution
through these phases is still subject to considerable uncertainties, in 
particular regarding  mass loss and the efficiency
of the third dredge-up. 
Stars within an intial mass range of $\sim$0.8 to 8\Msun ~end as WDs with 
a narrow mass range, mostly below 0.8\Msun.
It is  important to understand 
how the mass of  their  precursors relates to the final WD mass, in order to understand  the
relative contribution of different stars to the chemical enrichment of 
elements such as He, C, N and O.

While the evolution of the WD progenitors in the main sequence 
phase is 
fairly well understood and observationally constrained,  the hot-WD population is hitherto 
quite elusive,
owing to their small radius, hence low optical luminosity, 
 and extremely hot temperatures, to which optical colors are insensitive 
  (see e.g. \cite{bia07},  \cite{bia07apj}, \cite{bia07b}) 
as well as to their very short lifetimes on the constant-luminosity post-AGB phase.
To make matters more difficult, the post-AGB luminosity at a given stellar temperature varies significantly according to
the stellar mass, making it impossible to infer absolute luminosity from other physical 
parameters. 
The evolutionary time spent on the constant-luminosity post-AGB
phase  and on the cooling track is also a strong function of the mass (Vassiliadis \& Woods 1994). 
Therefore, the exact relation between progenitor's initial mass and WD mass (initial-final
mass relation, IFMR), remains to date
a crucial missing link in our understanding of stellar evolution and chemical enrichment
of the Galaxy. 

A characterization of the population of hot WDs in the Milky Way can reduce these uncertainties, 
and lead to a better understanding of processes that drive the chemical evolution of galaxies like the Milky Way. 
 UV photometry combined with optical measurements significantly increases the sensitivity to the
hottest temperatures. For example, the color difference between a \teff=50,000K and 20,000K star is
about 1.5mag in FUV-{\it g}, but  $<$0.4mag in U-B, and $<$0.15~mag in {\it g-r} which are comparable to photometric
errors when large surveys are considered. The sensitivity gained by extending the measurements to UV wavelengths is more critical  
for discerning the hottest stars (see e.g. Fig.s 5-7 of \cite{bia07apj} and Bianchi 2009).

The census and characterization of cool compact objects has significantly improved 
in recent years thanks to optical and
IR surveys.
The \cite{eisenstein06} catalog from the SDSS fourth data release (DR4) contains 9316 spectroscopically confirmed
WDs and 928 subdwarfs over an area of 4783deg$^2$; about one fourth (2741) have \teff$>$18,000K as estimated by the SDSS 
pipeline automated spectral analysis. An additional $\sim$5-6000 WD are expected from DR7 (\cite{knk09}).
Seven ultracool WD were added to the census by \cite{harrisetal08}.
Gontcharov et al (2010) extracted from 2MASS, Tycho-2, XPM and UCAC3 catalogs combined 
34~WDs, 1996 evolved (11,000$<$\teff$<$60,000K) and 7769 unevolved (\teff$<$7,000K) subdwarfs, using multicolor
photometry and proper motions of stars with 6$<$K${_s}$$<$14~mag. 
The current version (2008) of the  McCook \& Sion (1999) catalog of spectroscopically confirmed WDs includes 10132 entries (all types),
a factor of five increase over a decade, with respect to the original (1999) version  listing 2249 WDs.
\cite{vennesetal02} give a catalog of 201 DA WDs, spectroscopically confirmed from the $\sim$1,000 H-rich DA
WDs discovered in the 2dF QSO redshift survey;  Croom et al. (2004) 2dF catalog includes  2071 WDs over 2000~deg$^2$. 
 WDs in binaries from the SDSS were catalogued and studied by several authors, see e.g. Rebassa-Mansergas et al.(2010)
and references therein, Silvestri et al (2007), Heller et al. (2009, WD-M~star binaries) from spectroscopy;
other authors used optical-IR photometry to search for binaries among the known WDs 
(e.g. Tremblay \& Bergeron 2007, Watcher et al. 2003). Special classes such as 
CVs are addressed by other works (e.g. Szkody et al. 2009 and references therein, Gaensike et al. 2009),  
while others  studied in detail  the very local population (e.g. Holberg et al (2002) discuss
 122 objects within 20~pc from the Sun, which they estimate
to be an 80\% complete sample within this distance, and which include some double-degenerate systems). 

Finding the hottest, smallest stars, however,  remained
a challenge prior to the GALEX UV sky surveys, which provide deep sensitivity and large area coverage. 
For example, 105 were found in the original EUVE whole-sky
survey, with small subsequent additions obtained by combining EUVE and Rosat: see e.g. Dupuis (2002)
for a review and discussion.  The Rosat whole-sky survey produced 175 WD in X-rays, mostly DAs (Fleming et al. 1996, from Rosat PSPC).

Such catalogs of confirmed or candidate WDs enable the study of these objects as astrophysical probes,
of stellar evolution, of MW structure, of the local neighborhood, 
etc. Most importantly, comprehensive catalogs enable selection of targets 
for follow-up spectroscopy, which then provides the detailed physics of these objects, 
especially when extended to the UV and far-UV wavelengths, as proven by the score of results
enabled by IUE, FUSE, and HST spectrographs. 

Bianchi (2009, 2007), Bianchi et al. (2009a, 2007a, 2005) have demonstrated the power of 
far-UV and near-UV measurements, afforded for the first time over large areas of
the sky by GALEX, to unambiguously detect and characterize extremely hot stellar sources.
Not only the UV wavelengths are more sensitive to the temperatures of hottest stars, but
the combination of UV and optical colors allows a better separation of different classes of
astrophysical objects (e.g. Bianchi 2009 and references therein).

Our present work provides a selection of hot-star candidates from UV photometry; 
about 40,000 hot-star candidates with FUV, NUV photometric errors$\leq$0.3mag
(about 74,000 with photometric errors$\leq$0.5mag)  have also 
SDSS optical photometry.
The majority are likley  hot
WD with log(g) between 7 and 9. The catalog  covers different latitudes and enables a
first analysis of this stellar population with Milky Way models. 
Larger catalogs of UV sources with and without optical SDSS match are also constructed,
and made available as online products. 

In Section \ref{s_data} we describe the method  used to construct  ``clean'' catalogs of unique UV sources
from two GALEX surveys with different depths and coverage, and  subsets of these catalogs  with matched optical 
photometry, and we present the catalogs' characteristics.
 In Section \ref{s_catalog} we extract samples of hot-star candidates,   
and we analyze them  with Milky Way models
in Section \ref{s_discussion}. Discussion  
 and conclusions are given in Section \ref{s_conclusions}. 

\begin{figure*} 
\includegraphics[width=120mm]{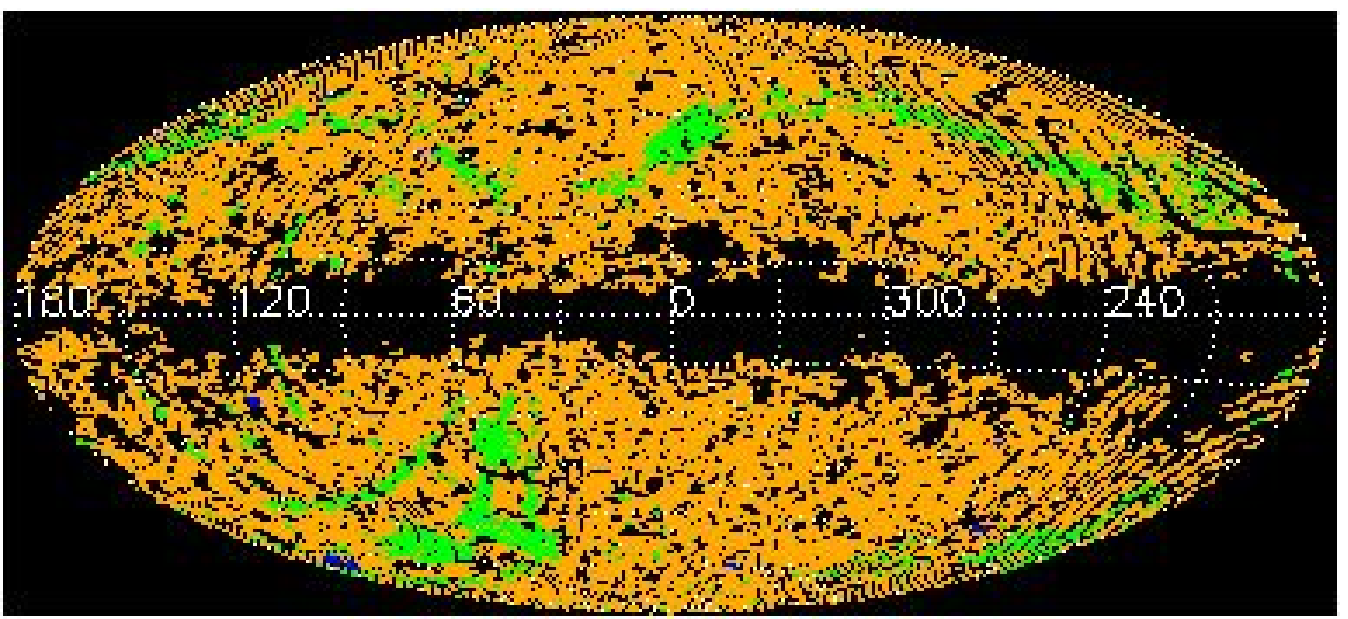}
\includegraphics[width=120mm]{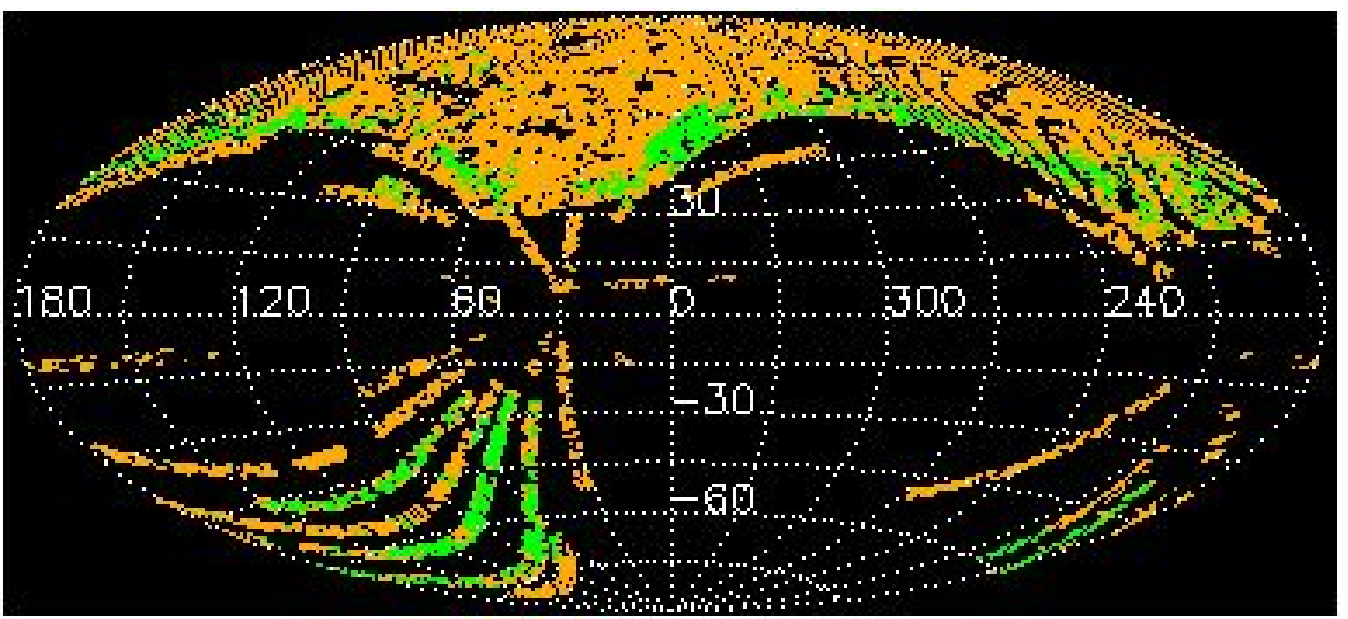}
\caption{TOP: Sky coverage (in Galactic coordinates) of the GALEX data release GR5 showing 
the major surveys: AIS (orange), MIS (green), DIS (blue).
BOTTOM: the portion of the GALEX GR5 sky coverage for AIS and MIS overlapping with SDSS seventh data realease  (DR7) }
\label{f_coverage}
\end{figure*}

\section{ The UV Sky Surveys and the source catalogs}
 \label{s_data}

\subsection{The sky survey data}
\label{s_datasurvey}
The Galaxy Evolution Explorer (GALEX) is imaging the sky in far-UV (FUV, 1344-1786\AA, $\lambda_{eff}$ = 1538.6\AA) 
and near-UV (NUV, 1771-2831\AA, $\lambda_{eff}$ = 2315.7\AA)
simultaneously, with a  field-of-view of 1.2$^{\circ}$ diameter  and a resolution of 4.2/5.3$''$
[FUV/NUV] (\cite{Morrissey07}).  The images are sampled with 1.5\as~ pixels.
Nested surveys with different depth and coverage are in progress. 
The widest sky coverage is provided by the All Sky Imaging Survey (AIS) and the
Medium [depth] Imaging Survey (MIS), that reach typical  depths of 
19.9/20.8mag (FUV/NUV) and 22.6/22.7mag (FUV/NUV) respectively, 
in the AB magnitude system.
The Nearby Galaxy Survey (NGS, Bianchi et al. 2003, Bianchi 2009, Gil de Paz et al. 2007), with  over 300
fields at MIS depth,  targeted nearby, hence fairly extended, galaxies, therefore it has been 
excluded in the present work, lest some sources from galaxies ``shredded'' by the pipeline  intrude on our catalog. 
See also \cite{bia09}, \cite{bia10}, \cite{biauvsky10}, \cite{bia07apj}, for a general discussion of the content of the UV sky surveys. 

In this paper we use data  from the GALEX fifth data release (GR5)
 AIS and MIS surveys, which include a total of 28,269 and 2,161 fields
respectively. The data are taken from the MAST archive. 
We restrict the catalogs to sources within the central 1$^{\circ}$ diameter 
of the field (for good photometry and astrometry, and to exclude edge artifacts).
With such restriction,  and eliminating overlaps, these
surveys cover a total unique area of 21,434.8 (AIS) and 1,578.6 (MIS) square degrees (Section \ref{s_area}).
Section \ref{s_catUV} describes the construction of the catalog of unique GALEX sources. 
In order to separate the UV sources by astrophysical classes, we examine in this work
the portions of the GALEX GR5 AIS and MIS surveys  that are also included in the 
 footprint of the Sloan Digital Sky Survey  seventh data release (DR7), which provides five optical
magnitudes: {\it u g r i z} in addition to the GALEX FUV, NUV magnitudes. 
The overlap between GALEX GR5 and SDSS DR7 includes 
 10,316~/~1,655~(AIS/MIS) GALEX fields, and the area coverage of the
overlap is   7,325 (AIS) and 1,103 (MIS) square degrees. Details of the area calculation are provided in Section \ref{s_area}. 
The sky coverage of AIS and MIS in the GALEX 
data release GR5 and its overlap  with SDSS DR7 are shown in Fig. \ref{f_coverage}.

\begin{figure*} 
\begin{center}
\includegraphics[width=130mm]{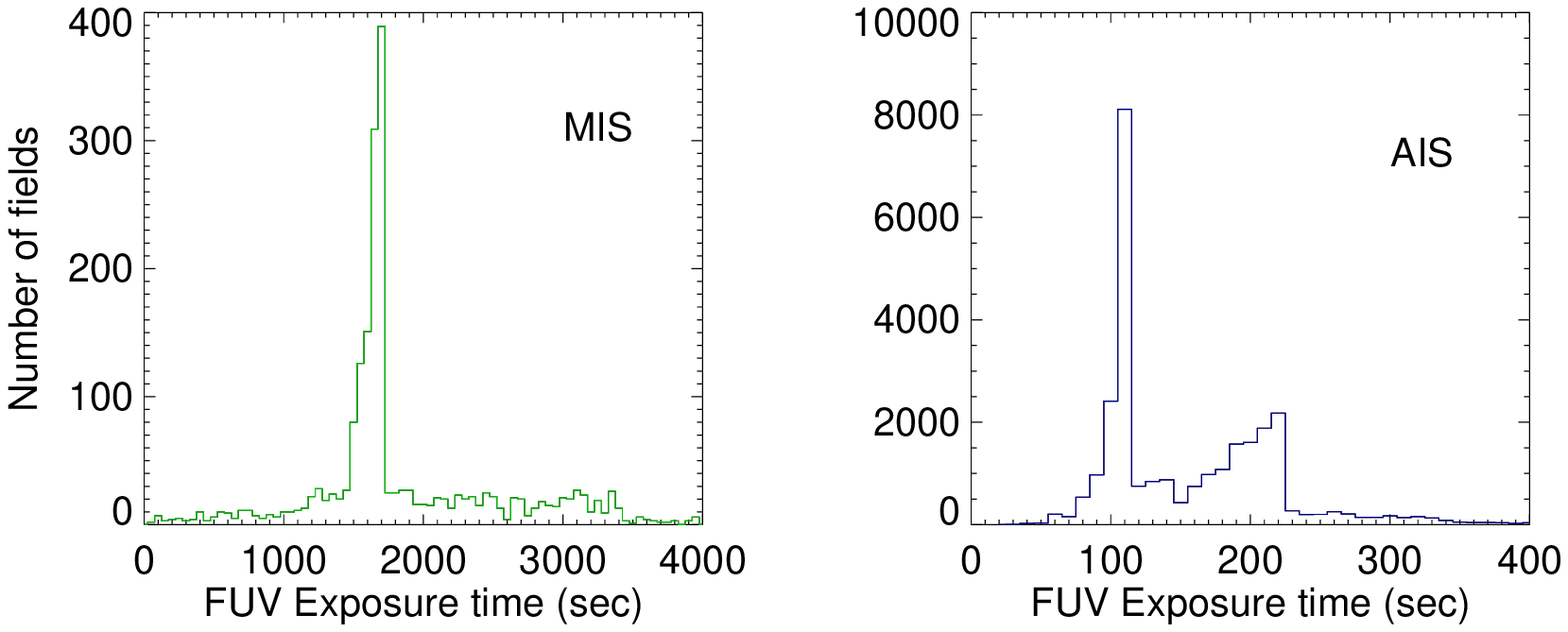}
\includegraphics[width=130mm]{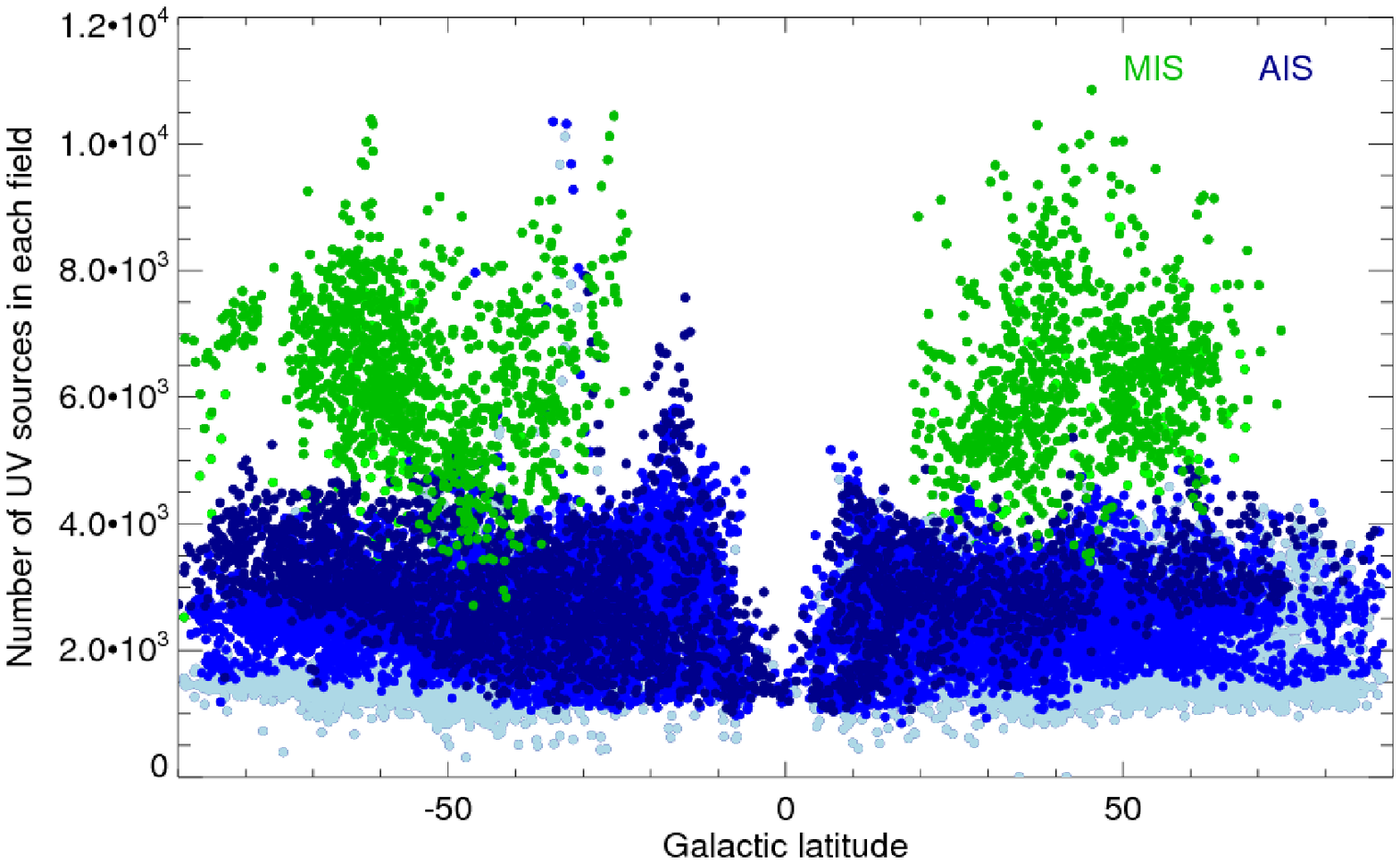}
\end{center} 
\caption{\small TOP: distribution of FUV exposure times  for AIS (10sec bins) and MIS surveys (50sec bins); a few fields have longer exposure times,
off the scale of the plots.
~~ BOTTOM: Number of UV sources (all UV detections, not just our selected hot stars)
in each  GALEX  field (1$^{\circ}$ diameter), before merging the catalogs and
removing overlaps. 
Three shades of blue (light/medium/dark) for AIS indicate exposure
times of $<$120~/~120-220~/~$>$220~sec respectively, and lighter~/~darker green for MIS fields indicate FUV exposures less/more 
than 1200~sec. 
Generally the number of UV source detections  in a field increases with depth
of exposure,  as expected, and as seen more distinctly for the AIS at the high latitudes, uncomplicated by dust
extinction. 
A few fields, including some with short exposures,  have overdensities. While the number of MW stars
increases towards the MW disk, as shown by the AIS sources, at MIS depth most UV sources are extragalactic, hence 
show no correlation with Galactic latitude except for the foreground extinction. 
The sharp drop in the center, due to dust extinction, defines the MW dust disk.   }  \label{f_exptime}
\end{figure*} 

\subsection{The catalog of unique UV sources}
\label{s_catUV}

Here we describe the procedure used to construct 
the catalog of unique GALEX sources (i.e. eliminating repeated measurements). 
 All catalogs described in this paper are
made publicly available from our website http://dolomiti.pha.jhu.edu/uvsky , 
and from MAST at http://galex.stsci.edu , and as High-Level Science Products (http://archive.stsci.edu/hlsp/ , in the ``Catalogs'' section). 
 Therefore, we provide here  information on how they were
constructed that will be relevant for potential users, as well as to others interested in
constructing future versions of similar samples. 

We extracted catalogs of GALEX sources from STScI MAST (www.mast.stsci.edu), at the CASJobs SQL interface
(www.galex.stsci.edu). Sources were extracted from the table ``photoobjall'',
from the MIS and AIS surveys separately, 
with the criteria that the source distance from the field center had to be  $\leq$0.5$^{\circ}$
and the photometric error less than 0.5~mag in NUV.
In other words, we initially included in our general  GALEX source catalog all
NUV reliable detections, regardless of whether they have also an FUV detection. 
For our selection of hot star candidates, we will eventually impose the  additional criterion of good FUV photometry. 
The additional restriction of err$_{FUV}$$<$0.5~mag significanly reduces the number of sources (by
a factor of up to 10, see Table \ref{t_stats}), with respect to the total number of NUV detections,  
 and of course introduces a bias in the
source catalog,  the hottest and bluest sources (the subject of this paper) being not affected but the redder sources
being progressively eliminated, as discussed in sections \ref{s_errcuts} and  \ref{s_stats}.
For more discussions about statistical properties, and biases inherent to sample selections, see \cite{biauvsky10}.

A few observations planned as part of  the MIS or AIS surveys actually 
partly failed, and resulted in one of the detectors (most often FUV) not being
exposed (such observations are typically repeated later). 
The observations with one of the two bands having zero exposure time
 were eliminated from our catalog, otherwise they will bias the statistics of FUV-detection
over number of total NUV detections, and the corresponding fields were not counted in the area calculation. 

Fig. \ref{f_exptime}  shows the distribution of exposure times (FUV is
shown, but NUV is generally equal or larger), for the AIS and MIS fields. The typical 
exposure time for MIS is 1500~sec, which is met or exceeded by the majority of fields. 
The AIS survey aims at exposure times of the order of $\sim$100~sec. We retained   
also fields with exposures shorter than typical.
Therefore, while the typical depth of the two surveys (AIS and MIS)  differ by $\sim$2~mag, the exposure
level is not strictly homogeneous across each catalog. 

  Fig. \ref{f_exptime}-bottom panel  shows  the number of GALEX sources in each GALEX field.
We color-coded the fields by three
ranges of exposure time for the AIS (two for MIS), since the number of sources detected 
above a given error-cut generally increases with exposure time. This plot is useful to check for
fields with overdensities, since the surveys, the broad AIS in particular, 
include also some stellar clusters.
In general 
the spread in number of sources per field is just about a factor of two for MIS. A few  AIS fields
around latitude -30$^{\circ}$ have an overdensity of almost one dex, however 
the total number of AIS fields in each 10$^{\circ}$-wide latitude bin (used in our analysis, see next section) is very large,
and a few overdensities do not affect our  analysis of  stellar counts 
with Milky Way models over wide areas. Fields with high density of sources, and in particular the stellar clusters
included in the surveys,  will be separately analyzed elsewhere. They are included in our
catalog for completeness and for possible use by others, although customized photometry is
desirable in very crowded fields (e.g. de Martino et al. 2008). 

 Nearby galaxies are generally observed as part of  the 'NGS' survey
which is excluded from our catalog, 
but a few large galaxies are also in the footprint of AIS and MIS and bright knots of galaxies 
 ``shredded'' by the pipeline may enter the catalog as separate sources (see \cite{bia07apj}). A few such sources may have 
FUV-NUV$<$-0.13 and appear pointlike in the SDSS catalog, and therefore may enter our catalog
of hot-star candidates (section \ref{s_catalog}), but they would not
affect the statistical results. For other, more specific uses of our catalog, they may be 
removed by checking against a list of nearby galaxies, as we did in \cite{bia07apj}. 
Using the ``child'' flags from the SDSS pipeline, that track deblended sources, 
proved not to be useful in indentifying and weeding out sources from ``shredded galaxies''.

The searches and download of GALEX sources with the above criteria were done 
using the java CASJobs command-line tool
(casjobs.jar).  One  problem we often encountered when using this tool to extract
and download query results from the database is that sometimes
the download pipe gets broken prematurely, and the results are truncated without 
any warning being given.  In order to verify that the output file
contained all the results, a separate count query was run on the SQL
server and compared to the number of output sources in  the downloaded file each time.
It was necessary to subdivide the search in small latitude strips 
due to various limits set by the CASjobs interface
(e.g. query length, output file size), as well as to avoid the frequent problem of the long searches being 
interrupted.  

The GALEX archive 
  contains multiple observations of the same
source, when some fields overlap or are repeated. 
Having all measurements of each source is useful for
variability studies (which will be addressed in a forthcoming work), 
and for chosing the best measurement when several are available.
For our present 
purpose we constructed from the total output a unique-source catalog,
in the following manner.  GALEX sources were considered possible duplicates
if they lied within 2.5\as~ of each other.  If two GALEX
sources were within this distance, but had the same
``photoextractid'' (i.e., they are both from the same observation) they
were both considered unique.  Otherwise, they were assumed to be
multiple observations of the same source.  We choose, to
represent the unique source, the measurement with the longest NUV
exposure time.  In the case of equal exposure times, 
the observation where the source was closer to the field center was chosen
(i.e. the source with the
smallest ``fov\_radius'' value from 
the ``photoobjall'' table), as photometric quality is usually better in the central part of the field.

\subsection{The matched UV-optical source catalog}
\label{s_catmatch}

A portion of the GALEX survey areas is in the footprint of the SDSS DR7 (see Fig.1 - bottom panel), 
and for the UV sources in this area we
constructed a catalog of matched optical sources. 
We  uploaded the coordinates of our GALEX unique sources  
into the SDSS SQL interface (version 3\_5\_16 rev~1.70, at casjobs.sdss.org) and queried for
SDSS matches against the ``photoprimary'' table (SDSS source catalog that includes only unique sources) using an inital  search radius of 4.2$''$.
A match radius of 3\as was eventually used in 
the final catalog, as we shall see.

As with the GALEX searches, the SDSS searches were done in an
automated way on 1$^{\circ}$-wide  Galactic latitude strips using the casjobs
command-line tool.  We found that there is currently a bug in the SDSS version of
this tool (v0.03) that causes the last character of a file to be dropped when
extracting query results.  We compensated for this by writing our
query to pad the end of the file with an additional character.

\subsubsection{Multiple matches.}

\begin{figure} 
\begin{center}
\includegraphics[width=90mm]{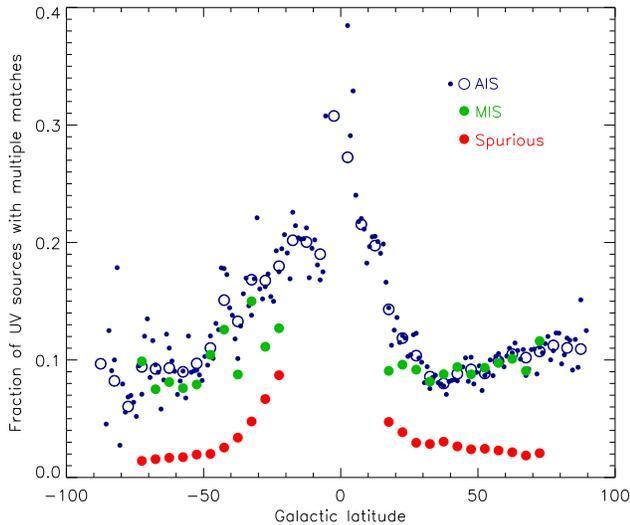}
\end{center}
\caption{Fraction of UV sources having multiple optical matches within a 3\as match radius,
as a function of Galactic latitude. Green dots show MIS data, blue dots show AIS data,
divided in strips of 1$^{\circ}$ (small dots), as well as averaged over 5$^{\circ}$ (blue circles). 
The increase towards the Galactic plane reflects the higher density of optical sources at low
MW latitudes. The plot also shows a slight North-South asymmetry. The fractions shown are  for point-like sources, with error cuts of 0.3~mag. 
The fraction increases if a larger match radius is used. The red dots are the incidence of spurious matches
(see text). 
\label{f_exptime2}}
\end{figure}

A given GALEX source may have multiple SDSS matches within the search radius, given the higher 
spatial resolution of the SDSS ($\sim$1.4\as). In such cases  the CASjob search
returns multiple lines with the same GALEX source and the various SDSS matches.  
We "ranked"  the multiple SDSS  matched sources based on the distance, with the
closest SDSS source being retained as the "match" for the GALEX source
and the additional SDSS sources being noted. 
When using UV-optical colors for source classification, UV sources with multiple
optical matches must be excluded, because, even though the closest optical source may be the 
actual counterpart, the UV photometry at the  GALEX spatial resolution of $\sim$5\as may 
 be a composite measure of two  optical sources,
and therefore the color would not be meaningful. 
The fraction of UV sources with multiple optical matches, eliminated from the analysis sample, is then 
taken into account when estimating density of astrophysical sources (\# per square degree).
This fraction is  given in Table \ref{t_stats}.    
Sources with multiple optical matches 
 are listed separately in our hot-star catalogs,
and are  included with ``rank''$>$0 in the total matched-source catalog. 
The relative number of UV sources with  multiple optical matches, shown in Figure \ref{f_exptime2},  increases towards low Galactic latitudes,
as expected.  The figure refers to our final catalog with match radius of 3$''$, and 
pointlike sources only: approximately 10\% of 
the UV sources have more than one optical counterpart, 
at intermediate and high Galactic latitudes, in agreement with our previous work
on earlier data releases which covered mainly high Galactic latitudes (Bianchi et al. 2005, 2007a). The fraction
increses to slightly over 20\% for latitudes $|$b$|$ $\approx$10-20$^{\circ}$, reflecting the higher density of stars
 in the MW disk, and is still very uncertain at  latitudes $|$b$|$$<$10$^{\circ}$, where we have 
little area coverage. For a larger match radius (4.2$''$), the fraction of multiple  matches
increases significantly, relative to  the total number of matches, 
and so does the incidence of spurious matches (Sect. \ref{s_spurious}), 
 therefore we adopted a final match radius of 3\as ~in our 
catalog.
The fraction is also higher if exended sources are included. 

\subsubsection{Other caveats.}

In the interest of others who may want to apply the same procedures, we also
mention  that, due to the large current  area coverage and consequent large 
number of sources, when it is necessary to perform the searches on small contiguous portions  of 
the sky (e.g. 1$^{\circ}$  strips), it may happen that the coordinates of a GALEX UV source
fall in one strip (close to the latitude limit), and the SDSS match falls in the next 
latitude range, due to a small difference in coordinates. 
Such cases are included in the final catalogs in the latitude range which
is appropriate according to the position of the GALEX source. 
 Care should also be taken when using the Galactic coordinates as returned by
the GALEX database, as those coordinates differ by $\la$0.5\as
from those derived by converting the GALEX RA, Dec coordinates to
Galactic coordinates using standard astronomical packages (e.g.,
WCSTools from the Smithsonian Astrophysical Observatory).  These
differences probably arise because the GALEX database stores 
Galactic coordinates using real data types, while using double
precision data types to store  RA and Dec values.  Therefore,
sorting GALEX search results based on the Galactic coordinates
returned by the database may give different results than using
WCSTools to derive those coordinates from the GALEX RA and Dec.

Another contingency that must be tested for, in the above procedures,  is whether a given SDSS source
matches multiple GALEX sources.  This can occur if a SDSS source lies
somewhere in  between two GALEX sources which were deemed unique.  In
this case, the match query  returns a match of each of the GALEX sources with the same SDSS object. 
We retain the match with 
the GALEX source which is the closest to the SDSS
source, and eliminate the other. This occurs rarely.

\subsubsection{Spurious matches.}
\label{s_spurious}

The probability of spurious matches was estimated as follows. 
 We randomly selected 30\% of
the coordinates from our matched GALEX-SDSS catalogs (in separate 5$^{\circ}$ bins
of Galatic latitude), and searched the SDSS database against 
 coordinates offset by 0.5$^{\prime}$ from those of the real sources.  
The spurious match rate, shown in Fig. \ref{f_exptime2} (red dots), is the
number of incidences where one or more matches were found (within a
search radius of 3") divided by the number of uploaded coordinates.
As for the multiple matches, it is higher  towards the Galactic plane, where the
stellar density increases and therefore  also the probability of random matches within a given radius. 
The rate  is of the order of a few\% at high and intermediate latitudes.
However, matches that are positional coincidences are likely to have odd colors
and the fraction may be lower in selected analysis samples. 

\subsubsection{Comparison with other  catalogs} 
We  point out, in the interest of users of our catalogs, 
that there are several differences from other basic matched source catalogs posted on MAST.
Most notably, we include only the central 1$^{\circ}$ diameter portion of the field,
therefore we conservatively  eliminate all edge sources (mostly defects or bad measurements
but also some good sources), but we retain all good portions.  Second, we  
consider all existing measurements for each source, from which the best is chosen. 
For example, the
  matched catalog  described by \cite{budavari09} instead, eliminates overlap by 
reducing each GALEX field to an hexagon (selection of primary). 
In this procedure, 
 the vertices of the hexagons include sources near the edge (i.e. possibly also sources with poor photometry, and artifacts).
Moreover, since the hexagons are fixed at nominal positions, but the actual pointings may
differ, unnecessary gaps between fields (which may not exist in the actual observations) are
introduced, as the authors point out.
Our partition, described in section \ref{s_area}, and following \cite{bia07b}, avoids such problems and
enables better and more homogeneous photometry quality, as well as easy and precise calculations
of the unique-area coverage for our GALEX surveys, and of overlap area with any other survey,
consistent with the actual source catalog rather than with nominal field centers.  This is preferable for our analysis.
Such differences may cause samples selected from  different matched catalogs, to differ slightly.

\subsection{Area coverage}
\label{s_area}

In order to derive  the density of sources extracted from our catalogs,
we  computed the total areas of unique coverage, taking into account  overlap between fields, 
in the GALEX GR5, and then the area  overlapping with SDSS DR7, following the method of \cite{bia07b}.
Since we restricted the catalogs to sources within the central 
1$^{\circ}$ of the GALEX field,  for each   
field we considered an effective radius of 
0.5$^{\circ}$.
Our  code  scans the entire sky 
and calculates the unique area covered 
by the GALEX fields, and the portion of this area covered also by the SDSS. 
First,  the whole sky is divided  in small, 
approximately square, tassels, along Galactic  longitude and latitude. 
We used steps in {\it l,b} of 0.05/0.1$^{\circ}$  for MIS/AIS respectively. Then, we 
find the distance between the center of each tassel in our whole-sky grid and the center of each GALEX  
field,  and sum the areas of the tassels that are within half degree
of a field center, avoiding to count the same tassel twice, which eliminates  overlap between fields.

The error in the estimated area  depends on the step used to calculate the grid of tassels (tassel size),
and the number of tassels. The area of each tassel decreases at high latitudes (the step  
in {\it l,b} is kept constant over the whole sky), hence the statistical error due to the tassels  along
 the field's edge (i.e. tassels that fall partly inside and partly outside the 1$^{\circ}$ circle\
of a GALEX field) 
varies with the location of the fields, both with latitude 
and with the relative location of the field center to a grid step. Such errors cancel out
statistically for a large number of fields. For a very small area however, i.e. for a few fields,
it is desirable to use smaller tassels.  
We estimate the uncertainty by computing the areas several times, each time offsetting the positions of the GALEX 
field centers by about half the size of a tassel, in both latitude and longitude directions.
The resulting uncertainty is $\la$1 square degree, 
for our total area and the areas of each latitude bin.

In order to calculate the area of overlap between the GALEX MIS and AIS surveys and SDSS DR7, 
we initially matched 
the centers of all tassels  included 
in our GALEX coverage  (MIS and AIS) to the SDSS DR7 footprint:  the areas of
tassels deemed within the SDSS footprint were summed to obtain  the area coverage of our matched source catalogs.
However, we  discovered an issue in the SDSS database DR7 footprint that prevented its use for calculating the  overlap areas. 
The output returned from the SDSS database footprint query is currently somewhat incorrect, perhaps due to some
SDSS fields which appear included in the  DR7 footprint in the database (therefore counting towards area coverage
in any estimate), however do not have  sources in
the "photoprimary" table or corresponding images because the observation failed or
was marked as bad for some reason.
We have consulted several experts of the SDSS database, but we found no way so far to identify 
such fields specifically. When we plotted on the sky the SDSS matched sources (obtained from the  SDSS ``photoprimary" catalog), and the 
sky-tassels deemed by the footprint-query to be within the footprint, we noted some not irrelevant areas of
mismatch (most notably, in the latitude strip 50-60$^{\circ}$South we found the largest ``false positive" area,
about 17.5deg$^2$ in total; the discrepancies are much smaller at other latitudes). 
We attempted to perform the footprint query on the SDSS database using different syntaxes. Specifically,
a basic all-inclusive footprint SQL search returns several ``false-positive" answers (i.e. several contiguous areas that the database considers
within the footprint, have no SDSS sources), hence the area based on such outputs would be overestimated with  respect to
the source catalog; a more detailed footprint query,  confining the search to "PRIMARY" footprint area, returns both
false-positives and false-negatives. 
We  finally could not use the SDSS footprint query, and 
 wrote  a separate code to  independently search,
for our entire GALEX sky coverage, sub-areas with and without SDSS sources; we considered the areas devoided of 
any SDSS sources as outside the SDSS footprint\footnote{These regions may be 'in' the footprint for SDSS database purposes,
but they contribute no sources to the catalog, hence counting them towards the area coverage would be inconsistent}. 
This procedure was complicated because (1) in order to decide whether a tassel contained SDSS sources, we had to consider SDSS sources of
any color, not just our hot-star catalog (which includes the rarest stars in the sky, hence some tassels may actually
not have any hot star even though they are located within the footprint), and (2) we had to consider much larger sky-tassels than our original     
grid, to avoid false-negatives, and then iterate within the "positive" tassels with a progressively finer 
grid (down to 0.05deg$^2$, about 1/16th of a GALEX field), to
confine the uncertainty of the area estimate.
It is important for our analysis, which compares density of sources (\#/square degree) with model predictions,
to estimate areas {\it consistent} with the source catalog used, in this case the PRIMARY catalog from the SDSS.
 We used our area estimates in the following analysis.
We stress, for future reference, that there was no way to detect the inconsistency between catalog and database footprint
other than by plotting the distribution of sources and footprint-tassels on sky coordinates in various ways.
This is a desirable test on any such work. 

The  GALEX GR5 AIS and MIS sky coverage, and the GALEX-SDSS overlap area in the matched GR5xDR7 releases,  restricting the GALEX field
 to 1$^{\circ}$ diameter and eliminating overlaps between GALEX fields, are given 
in Table \ref{t_area}, for the whole coverage and for separate ranges of Galactic latitude.

\subsection{Catalog statistics and effect of error cuts on the sample} 
\label{s_errcuts}

\begin{figure*}  
\begin{center}
\includegraphics[width=85mm]{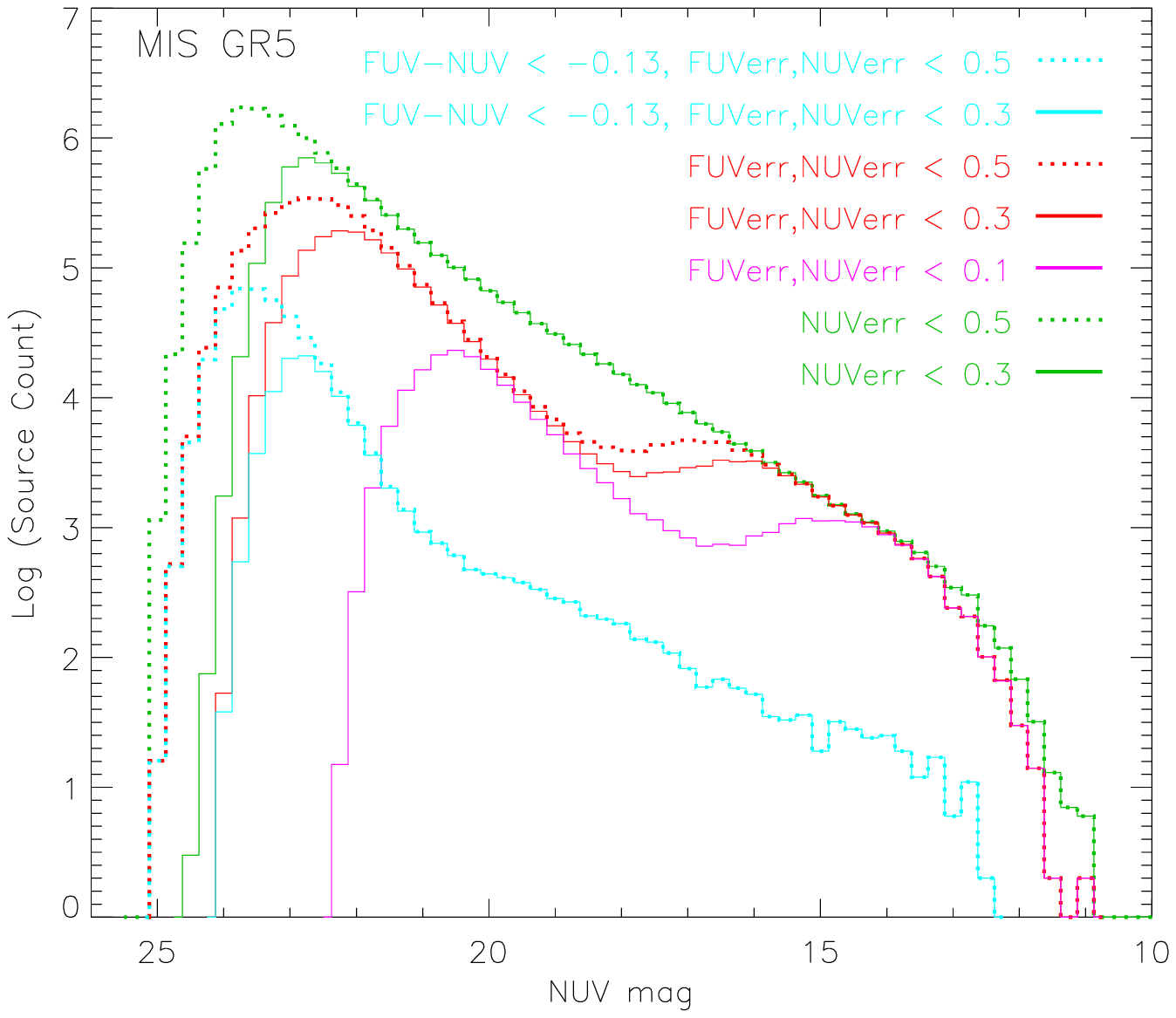} 
\includegraphics[width=85mm]{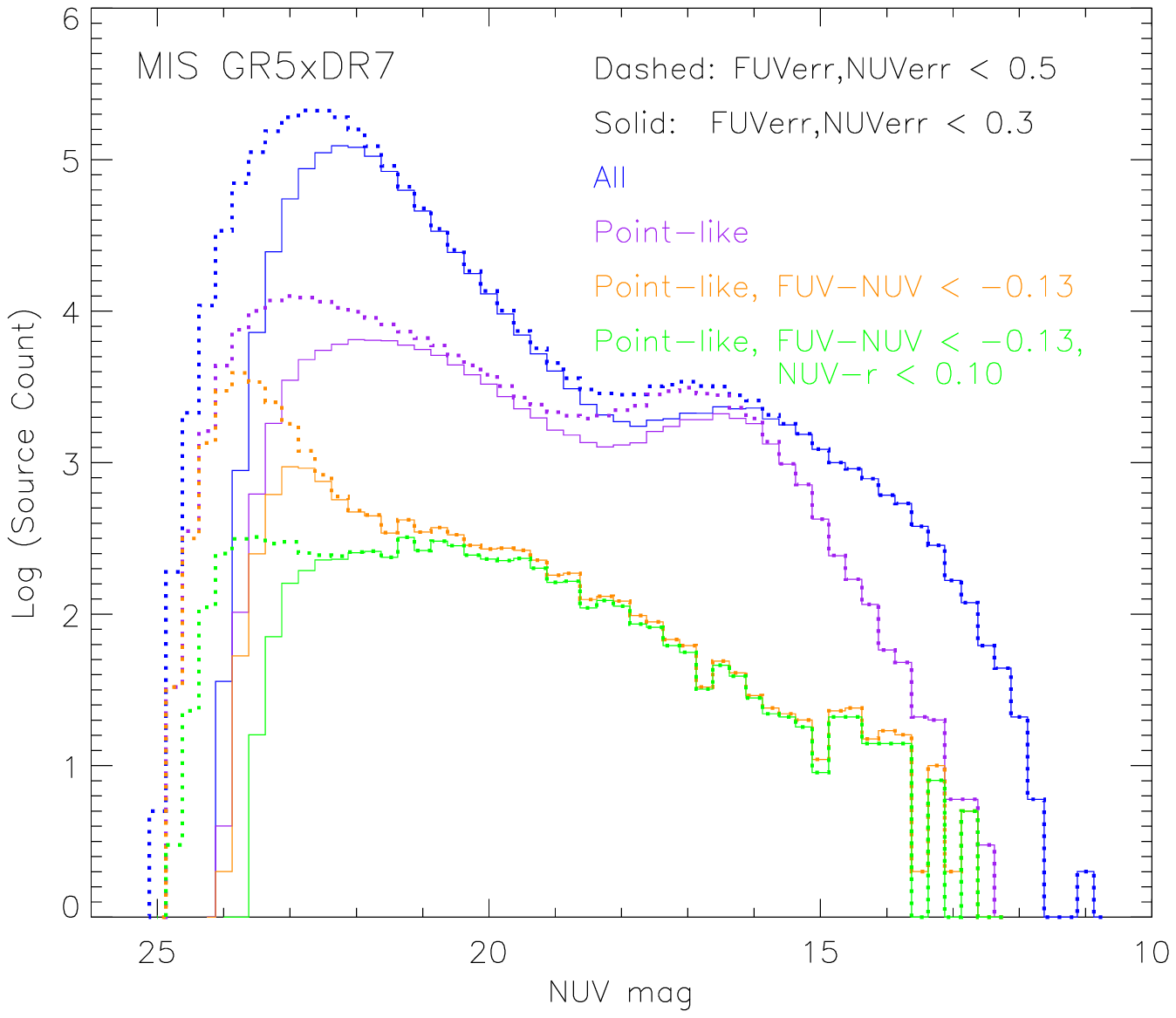} 
\end{center}
\caption{\small Distribution of UV sources in NUV magnitudes, and effects of error cuts
and color cuts. {\bf LEFT}: the whole catalog of GALEX unique sources (MIS), described in Section \ref{s_catUV},
 with progressive error cuts 
in NUV, and in both FUV and NUV.  
A more stringent NUV error cut 
simply causes a brighter cutoff at the faint end,
as expected, while the requirement of significant detection (or good photometry) also in  FUV 
modifies the overall distribution (``redder'' sources being eliminated), which becomes double-peaked.
The distribution of the ``bluest'' sources (FUV-NUV $<$-0.13), which include hot stars (this paper) and some QSOs 
(Bianchi et al. 2009a) 
is also shown.  {\bf RIGHT:} NUV magnitude distribution of the matched UV-optical sources. Here we also use the 
spatial information from the optical data, to separate point-sources (eliminating most galaxies,
but not QSOs with a high contrast between the central source and the underlying galaxy).
The point-like sources with FUV-NUV$<$-0.13 and FUV, NUV error  $\leq$0.3mag (solid orange line)
are the hot star candidate sample analyzed in this  paper, and the green histogram are the ``single'' hot stars. 
The figure, with the analysis in  section \ref{s_discussion},
shows that a potential halo component (NUV$\sim$24-25mag) is eliminated by our error cut at the current 
survey's depth; on the other hand, including objects with  larger errors would 
cause our color selection to include  low z QSOs.
}
\label{f_stats}
\end{figure*}

Figure \ref{f_stats} shows the UV~magnitude  distribution of our  GALEX unique-source catalog,
and the effect of progressive error cuts in NUV, and in both NUV and FUV. 
While a progressively stringent error cut in one band (NUV) simply truncates the sample to a brighter magnitude limit, 
 an additional  error cut in FUV,
which effectively raises the faint limit of the sample in this band, causes a relatively
higher decrease of redder sources, and the histogram of sources distribution becomes double-peaked. 
The same effects are seen (right panel) for the matched GALEX-SDSS sources catalog. This catalog can also be
separated in pointlike and extended sources using the SDSS spatial information ($\sim$1.4\as~ resolution). 
Such distinction shows that most of the faint-magnitude peak of the source distribution is due
to extended sources, which are likely galaxies, as can be expected (e.g. Bianchi 2009). 
The most restricted sample in the right-panel histograms, the point-like matched sources with error $\leq$0.3mag
in both FUV and NUV, and FUV-NUV$<$-0.13, is the subject of this paper; it contains mostly hot
star candidates (\teff $>$ approximately 18000K), with some contamination by QSOs at faint
magnitudes and red optical colors, discussed in section \ref{s_purity} (see also \cite{bia09qso}).

\begin{figure*} 
\begin{center}
\includegraphics[width=80mm]{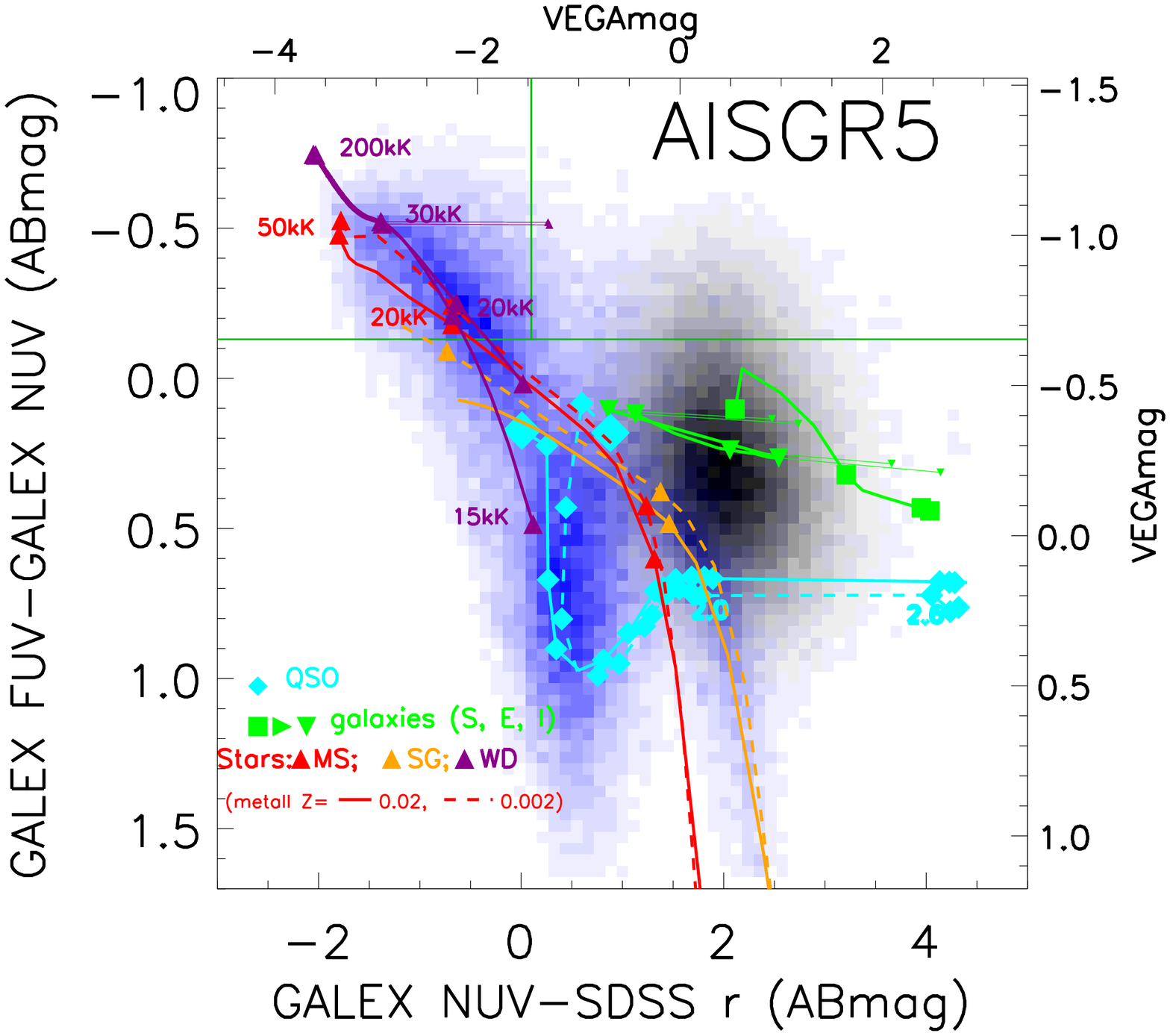}
\includegraphics[width=80mm]{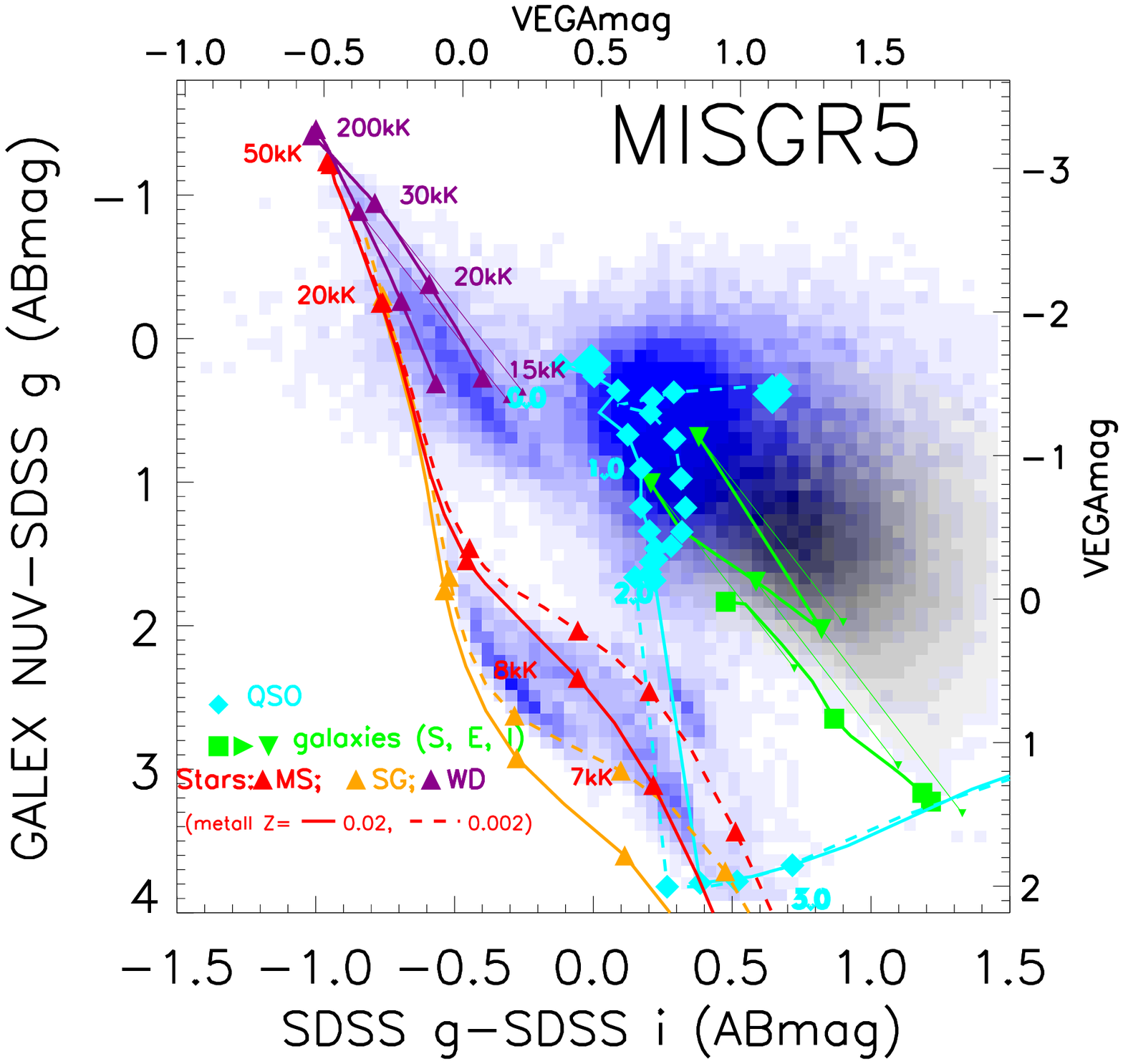}
\end{center}
\caption{Color-color diagrams for the GALEX UV sources matched to SDSS sources, at AIS and  MIS depth. Blue/black densities
are pointlike/extended sources respectively. Model colors for stars (T$_{eff}$ sequence, for
different gravities), QSO (redshift sequence, cyan) and galaxies (age sequences, green) are
shown. The two purple stellar sequences (label ``WD'') in the right-side panel are for logg=9.0 and 7.0.  
Although  gravity cannot be separated well photometrically at high \teff's, the vast majority of hot stellar
sources are clustered in between these two model sequences, and much fewer are seen along the logg=5 and 3  (red and yellow) sequences
at hot temperatures, as expected. 
In the $\approx$ 2.5mag shallower AIS survey the relative number of stars and
extragalactic objects is comparable, while at MIS depth the latter dominate. 
From such diagrams the hot star samples are selected (see text), with color cuts shown by the green lines in the left-hand plot.  \label{f_colcol} }
\end{figure*}

The analysis of the hot-star sample with Milky Way models (see later) shows that the  magnitude
limit introduced by our error cut of $\leq$0.3~mag eliminates mostly halo and thick-disk MW stars from the sample
(between the green solid- and dashed-lines on the right-hand panel).
On the other hand, including sources with larger photometric errors would introduce significant
contamination of the sample by non-stellar objects (see Fig \ref{f_colcol}). 

More statistical analysis of the catalogs, and discussion of potential biases in flux-limited sample
selection, are given by \cite{biauvsky10}.

\section{Selection of hot stars}
\label{s_catalog}

The matched GALEX-SDSS sources in our catalogs 
are shown in  color-color diagrams in Figure \ref{f_colcol} as density plots.
Model colors for major classes of astrophysical objects are also shown to guide the eye
in interpreting the distribution of sources. We refer to \cite{bia09qso} and references therein for other similar figures
and description of the model colors. 

We selected hot stellar candidates by extracting the matched pointlike sources with FUV-NUV $<$~-0.13
(in the AB magnitude system), which corresponds to stars hotter than 
$\approx$18,000~K, the precise \teff depending on the stellar gravity, and on the model atmospheres adopted (discussed later).\footnote{The measured colors also depend on extinction, however for a Milky Way-type dust,
with R$_V$=3.1, and moderate reddening amounts, the GALEX FUV-NUV color is basically reddening-free.
In fact, the higher selective extinction A$_{\lambda}$/\ebv~ in the FUV range is approximately equalled by the effect of the 
broad 2175\AA ~ feature in the NUV range, see Fig. 2, and Bianchi (2009). The absorption in each UV band, however, is
much higher than in optical bands. }
This FUV-NUV limit was chosen to largely exclude all extragalactic objects, based
on colors derived from canonical templates of QSOs and galaxies (e.g. Bianchi 2009 and  references therein, see also Fig. \ref{f_colcol}).

As shown by  Figure \ref{f_colcol} in this work, and by Fig. 7 (lower panels)  \cite{bia07apj}, at 
AIS depth ($\sim$21th mag)  sources with FUV-NUV $<$~-0.13 have very little contamination  by
extragalactic objects, but at fainter magnitudes 
a number of QSOs and galaxies with extremely ``blue'' FUV-NUV colors is found.
\cite{bia09qso} 
have examined the nature of such uncommon QSOs, and their Figure 1 and 2
show the location of these objects in the color-color diagrams. These QSOs 
have extremely ``blue'' FUV-NUV (more negative than typical QSOs at any redshift), 
but optical colors typical of QSOs and galaxies. Therefore they overlap,
in  UV--optical color-color diagrams, 
with the locus of hot stars with a cool companion, i.e. hot stars 
having very blue (i.e. negative) FUV-NUV color but optical colors redder than what their FUV-NUV would imply for a 
single hot source. We discuss the QSO contamination in section \ref{s_purity}.

We restricted the catalog to sources
with photometric errors $\leq0.3$mag in both FUV and NUV. 
 Fig. \ref{f_stats}, previously discussed, and Table \ref{t_stats} 
in this work show the effect of the error cuts on the sample's statistics.
We additionally use one optical band ({\it r}) to separate the hot sources in ``single'' star candidates,
having color NUV-{\it r}$<$0.1, and ``binary'' candidates having  NUV-{\it r}$>$0.1.
In more detail, the NUV-{\it r} boundary between single and binary hot stars has a complex dependence
on stellar \teff and radii of the binary pair, which we will address in a different work. 
For example, a
  hot WD with an A-type companion,  
and pairs with small mass ratios will fall in the color selection of our ``single'' hot-star sample. 
The adopted value is a useful cut 
to eliminate  most extragalactic sources contaminating our FUV-NUV$<$-0.13 sample, as they
have NUV-{\it r}$>$0.1, however 
it is obviously an oversimplification for a detailed study of binaries. In sum, our so-called 
(for simplicity) ``binary'' sample includes stellar pairs with 
a hot WD and a  cooler star, and some QSOs;  the ``single'' sample includes all single stars and some binaries.  
We impose no error cut on the {\it r} magnitudes, in order to not limit the 
sample of the hottest WDs, which are faint at optical wavelengths. 
If the restriction of error$\leq$0.3~mag  were imposed to the {\it r-}band, the density of objects in the
 MIS sample would be reduced by about 30\% (GALEX is
`deeper' than SDSS for very hot stars of low luminosity, see section \ref{s_additions}); the loss 
would be  much smaller for the brighter AIS sample.

The density of hot-star candidates is shown in Fig. \ref{f_lat} and will be discussed later.
The magnitude distribution of 
the total hot-star candidate sample (FUV-NUV $<$~-0.13) is shown with shadowed histograms in Figures \ref{f_strip_few} and beyond,
and the ``single'' hot star candidates (FUV-NUV $<$~-0.13 and  NUV-{\it r}$<$0.1) with solid-color  histograms.  

\subsection{Hot Stars not detected by SDSS}
\label{s_additions}
 
While the SDSS depth of 22.3/23.3/23.1/22.3/20.8~AB~mag in {\it u,g,r,i,z}  provides a fairly
complete match to the AIS UV source catalog, the hottest, smallest stars detected in
the MIS may fall below the SDSS limit.  This can be guessed by comparing 
in Fig.s \ref{f_strip_few}, \ref{f_alllat}, \ref{f_thind} 
the green-dashed histogram (matched GALEX-SDSS sources with FUV-NUV$<$-0.13)
with the light-green filled-color histogram 
(matched  sources with FUV-NUV$<$-0.13 and NUV-{\it r}$<$0.1, i.e. ``single'' hot stars).
The hottest single WDs will be   faintest  at optical wavelengths (e.g. a star with log~g=9. and 
\teff=50,000/100,000~K would have  FUV-{\it r}=-2.33/-2.61 in AB~mag, according to our
TLUSTY model grids). Therefore,  hot WDs still detectable in UV at the depth of  our MIS sample 
(see  Fig. \ref{f_stats}) may be below the SDSS detection limit, while hot WDs with an
optically-brighter binary companion or QSOs, which have redder UV-optical colors, will
be detectable also in the SDSS imaging.  This explains why the density of  MIS ``single'' hot matched
sources (light-green) drops at faint magnitudes earlier than the dashed-green histogram,
the difference between the two being larger than what can be ascribed to QSO contamination
 (Section \ref{s_purity}).  

 In order to estimate the incompletess  of faint hot-WD  counts in our matched sample, 
we searched for GALEX MIS sources with FUV-NUV$<$-0.13 that are within the SDSS DR7 footprint,
but do not have SDSS counterpart. 
 We found $\approx$1,500 such sources, 
having no optical match in our catalog with a match radius of 3$''$ (section \ref{s_catmatch}).
We performed a number of tests to verify if these are real sources. We matched them against the SDSS catalog, 
extending the match radius to 6$''$, and found 399 additional matches, 138 of them are classified as ``pointlike'' 
sources at the sdss resolution. These mostly appear to be  actual sources in the images, although 
we note that for faint sources, the SDSS classification of ``extended'' and ``pointlike'' is not always 
reliable, as shown by \cite{bia09qso}. 
It is expected that a small number of sources may have optical coordinates differing by more than 3$''$  from the UV
position; sometimes this is due to a nearby source not fully resolved.    
 For the remaining 1,100 UV sources with no SDSS match out to 6$''$, visual inspection of random subsamples revealed some to be
part of extended cirrus emission, many seem likely sources, a few cases are parts of a shredded galaxy. 
The `kron radius' from the GALEX pipeline gives an indication of the spatial extent of the source:
 189 sources have kron radius larger than 3.5, hence are
probably not stellar sources, but visual inspection suggests about 75\% of them to be real sources. 
Even visual inspection however is not always conclusive, for  faint sources or complicated fields,
and better resolution or deeper exposures would be needed for a final sample.    
In sum, a very large fraction of the 1,500 objects unmatched within 3$''$ are actual sources,
and an undetermined fraction may be actual hot stars.  

In spite of the large uncertainties, we added all
these objects to the ``single'' hot matched sources catalog, and show the total 
 as filled-dark-green histograms in Fig.s \ref{f_lat}, \ref{f_strip_few}, \ref{f_alllat}, \ref{f_thind}.
They must be considered a very generous upper limit, 
and only a reminder that
the light-green histograms suffer from incompleteness for
hot stars with  NUV fainter than $\sim$22mag.

 \subsection{Purity and completeness of the sample}
\label{s_purity}

In order to estimate the probability that the photometrically-selected hot star candidates  actually are hot stars, 
and the possible contamination of the sample by other types of objects, we examined the subsamples
of our catalogs  for which SDSS spectra exist.
Out of 9,028 MIS matched sources with FUV-NUV$<$-0.13 and  FUV,NUV error $\leq$0.3mag, 810 have SDSS spectra,
104 of which  are  sources with NUV-{\it r}$>$0.1, i.e. in the  ``binaries'' {\it locus}.
Of these ``binary'' candidates with existing spectra, 58 are spectroscopically classified 
by the SDSS pipeline as stars, 4 as galaxies, 42 as QSOs. 
Therefore, almost half of the MIS sources with  FUV-NUV$<$-0.13 and NUV-{\it r}$>$0.1 could
be QSOs (as found also by \cite{bia07apj}), and half could be hot stars with a cool companion. 
In more detail,  QSOs numbers  increase at fainter magnitudes, and have FUV-NUV colors closer
to our limit (-0.13) than the hot WD (\cite{bia09qso}). 
Because stars and QSO counts vary with magnitude in different ways, 
and especially because this statistic can be highly biased by the SDSS selection of spectroscopic targets, 
which is of course unrelated to  our present selection, 
and because the spectroscopic survey does not reach the depth of the MIS photometric survey,
we refrain from assuming a correction for the fraction of extragalactic objects in the binaries sample based on current data.

A much higher purity 
is found for the ``single'' hot-star candidates 
(FUV-NUV$<$-0.13 and NUV-r$<$0.1). Out of 4,924  in the MIS sample, 706 have spectra, 
703 of which are classified as stars, and two as galaxies. Only one is classified as QSO
by the SDSS spectroscopic pipeline, but it is actually a hot WD, as shown by \cite{bia09qso}.
  This implies a purity of almost 100\% for the ``single'' hot-star candidates, down to 
the magnitude limit of the SDSS spectroscopy at least. Again we stress however that the
spectroscopic subset is serendipitous for our purpose, but not necessarily unbiased.

In the AIS sample, out of  28,319 total hot-star candidates 
(21,606 of which with NUV-r$<$0.1), 
4448 (3737) have spectra, 
classified as 4075 (3721) stars,  309 (9) QSOs, and 59 (7) galaxies;
 corresponding to 91.6\% (99.6\% for ``single'' hot-star candidates) purity. 
The higher content of stellar sources in the AIS sample, even for the ``binaries'',  is due to the brighter magnitude limit.  

A crude, more direct indication that there is contamination by extragalactic sources in the
``binary'' hot-star {\it locus}, is that the fraction of ``binaries'' (NUV-{\it r}$>$0.1) among the
hot sources is about 25\% for the AIS and  45\% for the MIS. 
We note that the AIS value of 25\% is not reflecting the fraction of actual binaries, 
because the pairs whose components have  similar \teff~ are included in the ``single'' color cut,
 and on the other hand, AIS ``binaries'' may also contain QSOs. 

In view of the  contamination by QSOs in the ``binary'' hot-star sample, 
in the following  analysis with Milky Way models we will consider  the counts of ``single'' hot-star candidates,
and assume a canonical binary fraction of 30\% in the models.

\subsection{Characteristics of the Hot-Star Candidate Catalog}
\label{s_stats}

\begin{figure} 
\begin{center}
\includegraphics[width=80mm]{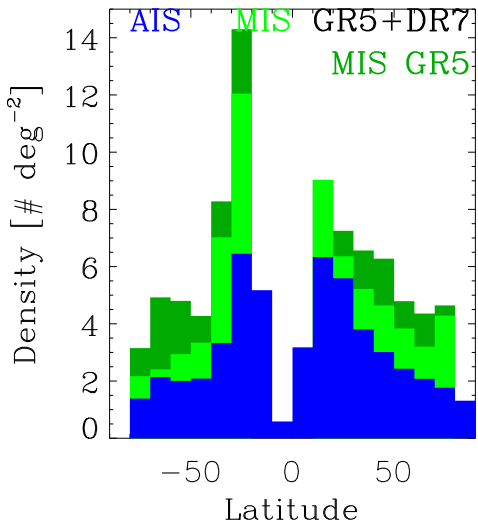}
\end{center}
\caption{Density of ``single'' hot stars 
 at different Galactic latitudes (all  magnitudes combined,
green for MIS, blue for AIS). The stellar counts are not corrected for Milky Way extinction,
which is more severe at low latitudes, and causes the sharp drop seen on the MW plane.
The bright-green histogram is the GALEX-SDSS matched MIS single hot-star candidates, the dark-green histogram
includes also hot 
UV sources with no optical match, within the SDSS footprint (see text). The ranges 10-20$^{\circ}$~N
and 70-80$^{\circ}$~N
have only 6 and 3~square degree MIS coverage in the matched footprint, therefore
the statistical significance is much less than in other bins.   \label{f_lat} }
\end{figure}

The number of hot sources selected  (pointlike, with error $\leq$0.3mag in  FUV and  NUV,  and  FUV-NUV$<$-0.13; see sect.\ref{s_catalog}), 
and of the subset with NUV-{\it r}$<$0.1 (``single''), 
are given
in the last two columns of Table \ref{t_stats}. 
Stellar counts as a function of magnitude will be analyzed in the following sections. 
Figure \ref{f_lat} shows the density of ``single'' hot-star candidates (all magnitudes combined) as a 
function of Galactic latitude. The density increases by over a factor of seven,
from the  poles towards the Galactic plane.
Owing to the MW extinction,  the counts are a lower limit at all latitudes, but especially near  the 
Galactic disk. Therefore,  the  variation with latitude  shown by this figure is less
than the actual one.  For the MIS, the density of hot stars in the matched sample (light-green histogram) 
is incomplete due to the 
SDSS limit (section \ref{s_additions}). The dark-green histogram shows the MIS ``single'' hot-star candidates 
including the hot GALEX sources without optical match.
Sources with NUV-{\it r}$>$0.1  are not shown due to the QSO contamination (section 4.3); adding the 
stellar binaries in this color range would increase the number density but not change the trend.

 In the next section we analyze the density of hot stellar candidates as a function of magnitude;
the samples are divided in strips of 10$^{\circ}$ Galactic latitude, in order to examine the  structure of this
 MW stellar component.

\section{Analysis. Comparison with Milky Way Models}
\label{s_discussion}

From the point of view of stellar evolution, the sample presented in
this paper includes essentially two kinds of evolved stellar objects: the
post-AGB stars that have just expelled their envelopes and are
crossing the HR diagram towards higher \teff's at constant luminosity, on their way to become planetary
nebule nuclei, and the hot white dwarfs which are fading both in
\Teff\ and luminosity. Post-AGB stars are very elusive because 
they evolve very fast, on time-scales of the order of
10$^3$-10$^5$~years (\cite{VW94}). Hot white dwarfs evolve
with longer timescales, but at significantly fainter
luminosities. Both are elusive at all wavelengths except the UV. Our
hot-star census based on the UV sky surveys provides the first
opportunity to examine a comprehensive, unbiased sample of such
objects.
Our hot post-AGB and WD candidate sample  extends over a significant sky coverage, and clearly presents a
disk-like distribution, concentrated towards the Galactic disk
(Fig.~\ref{f_lat}). Such sample will be analyzed in the context
of a model for the Milky Way geometry below. 

\subsection{Constraining Milky Way models}
\label{s_mwmodels}

In this section we analyze the number counts and sky distribution of our hot star 
candidates,  using the TRILEGAL stellar population synthesis code
(Girardi et al. 2005). TRILEGAL creates mock catalogues of stars
belonging to the Milky Way, and then distributes them along the
line-of-sight. It extracts the simulated stars from extended libraries
of evolutionary tracks and synthetic stellar spectra, assuming
reasonable prescriptions for the distributions of ages, masses and
metallicities of the Milky Way stellar components: thin and thick
disks, halo and bulge. We note here that the bulge component does not contribute almost
any source to our hot star sample, and is therefore not shown in the plots of  our model results.

The thin disk density, $\rho_{\rm thin}$, decreases exponentially with
the galactocentric radius, $R$, and as a squared hyperbolic secant
function in the direction perpendicular to the plane, $z$:
\begin{equation}
\rho_{\rm disk} = C \exp(-R/h_R) \, {\rm sech}^2(2\,z/h_z) \,.
\end{equation}
The scale length is set to $h_R=2800$~pc, whereas the scale heigth increases
with the population age $t$ as
\begin{equation}
h_{z}(t) = 95\,\left(1+t/(4.4{\rm Gyr})\right)^{1.67}    \,,
\end{equation}
Such an increase is necessary to describe the observed increase of the
velocity dispersion W  with age (see e.g. \cite{Holmberg_etal09}). 

The thick disk follows the same functional form but with both scale
parameters fixed as $h_R=2800$~pc and $h_z=800$~pc. The constants $C$,
for the thin and thick disks, are adjusted so that the surface thin
disk density is equal to  59~$M_\odot\,{\rm pc}^{-2}$, whereas the local
thick disk density is of 0.0015~$M_\odot\,{\rm pc}^{-3}$. 
The halo is modeled as an oblate spheroid following a deprojected
$r^{1/4}$ law, with a local density of 0.00015~$M_\odot\,{\rm
pc}^{-3}$. Finally, the Sun is located at $R=8700$~pc, $z=24.2$~pc.

The default parameters for the geometry of the  MW components  in TRILEGAL are
calibrated  to reproduce the star counts in a local sample
extracted from the Hipparcos catalogue, and in several multiband
catalogues including  the shallow, all-sky 2MASS, and  a few deep
surveys such as EIS-deep and DMS (\cite{Leo05}). 
For regions out of the Galactic plane,
i.e. for $|b|>10^\circ$, errors in the star counts predicted by
TRILEGAL are typically of about 10 to 20\% down to $K\sim14$ (Girardi
et al., in preparation).

The reddening is taken into account in the Milky Way models as follows. 
Along  a given line-of-sight, the value of $E(B-V)$ at infinity is taken
from the \cite{Schlegel_etal98} 
maps. The total extinction 
is then distributed along the line-of-sight as if it
were generated by an exponential dust layer with scale height
$h_{z}^{\rm dust}$; its default value is $h_{z}^{\rm dust}=110$~pc. In
this way, the closest simulated stars are unreddened, whereas those at
distances of a few hundred parsecs are reddened by the full amount
predicted by the \citet{Schlegel_etal98} maps.

\begin{figure}
\begin{center}
\includegraphics[width=80mm]{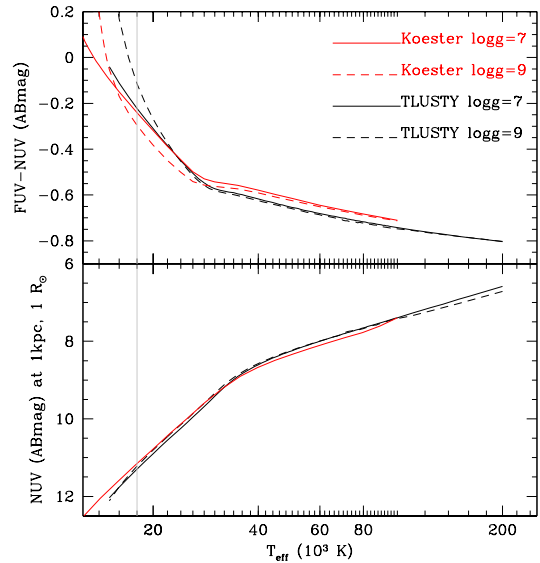} 
\caption{\small Comparison of model magnitudes (NUV) and color (FUV-NUV) of stars with log(g)=7.0
and 9.0, of varying \teff. Model magnitudes are  constructed from our grid of pure-H TLUSTY NLTE model
spectra  and from \citet{Koester08} LTE models. The latter include convection, which TLUSTY (ver198) does 
not. Our color cut of FUV-NUV$<$-0.13 
corresponds to \teff  $\approx$ 18,800/17,200K (log(g)=9.0/7.0) 
in the TLUSTY models, and to \teff=18,000K in solar Kurucz-model colors with log(g)=5.0 (not shown). 
While TLUSTY and Koester  model colors with log(g)=7.0 agree quite well for \teff up to 26,000K, 
the two grids are discrepant by $\sim$0.15mag for log(g)=9.0, where the Ly$\alpha$ wings are broad enough to enter the GALEX FUV band
when convection is taken into account. 
The situation reverses at hotter \teff 's. 
A color cut at FUV-NUV$<$-0.5 benefits from less scatter among model grids, which we could  take as a 
measure of less uncertainty, however the resulting sample (the very hottest stars) would be drastically
reduced, and so the statistical significance.  \label{f_models_FUV_NUV} }
\end{center}
\end{figure}

\begin{figure*} 
\begin{center}
\includegraphics[height=45mm]{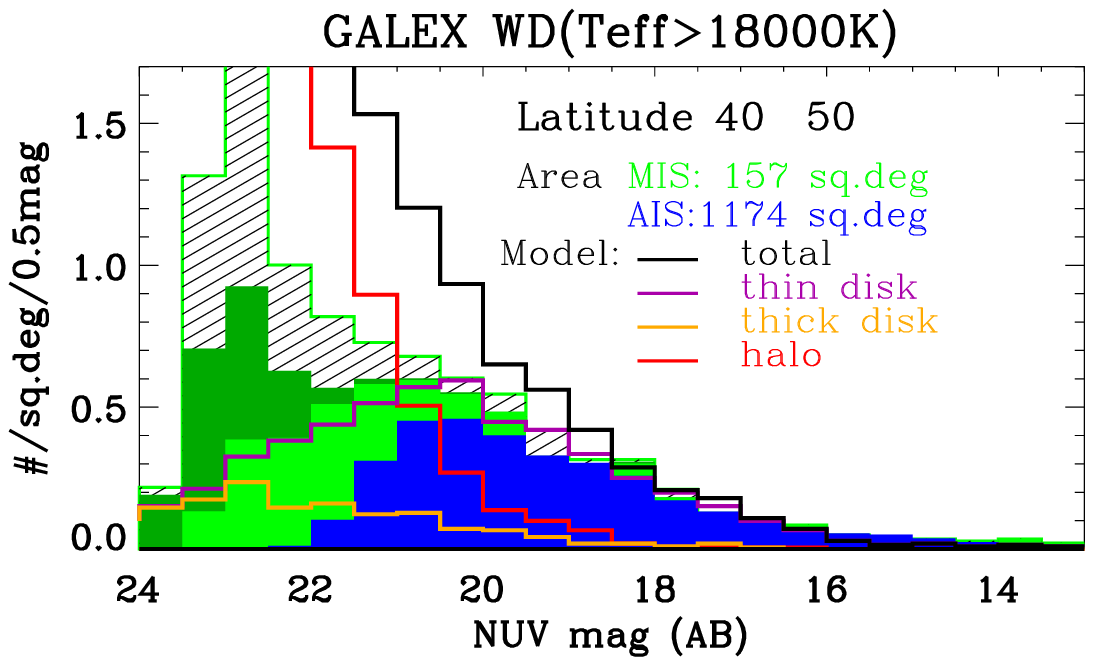}
\includegraphics[height=45mm]{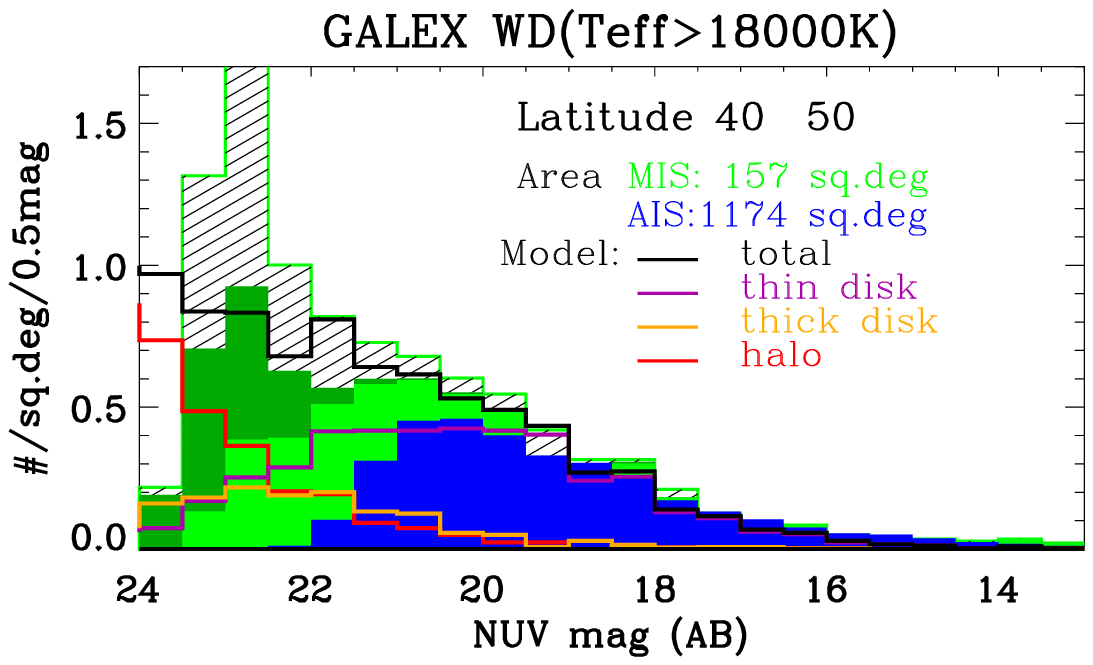}
\includegraphics[height=45mm]{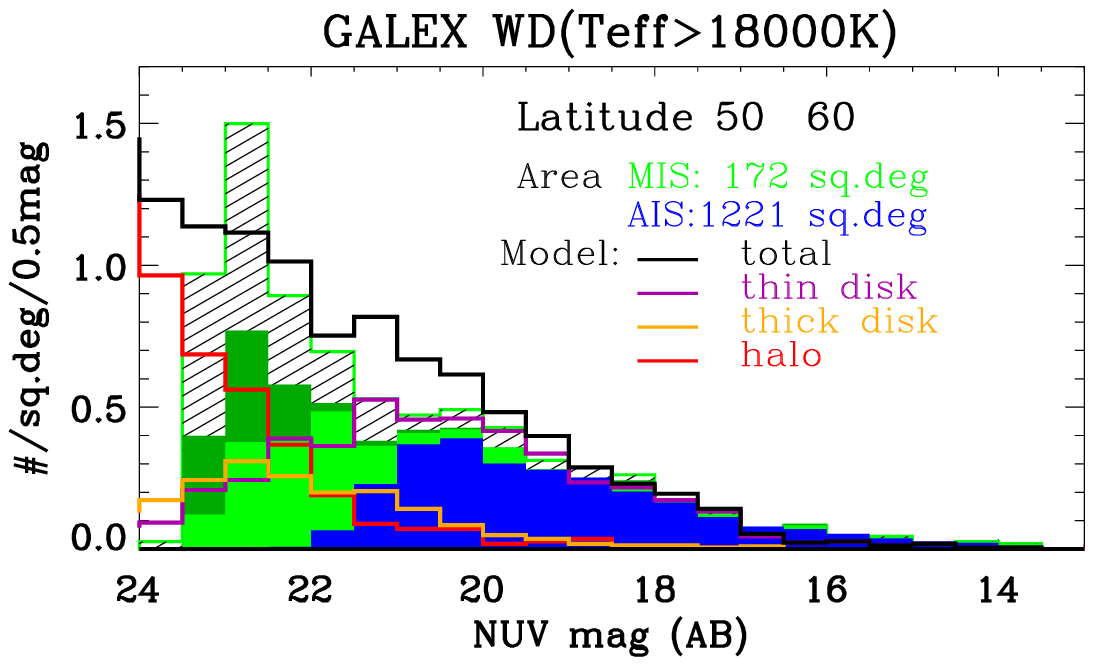}
\includegraphics[height=45mm]{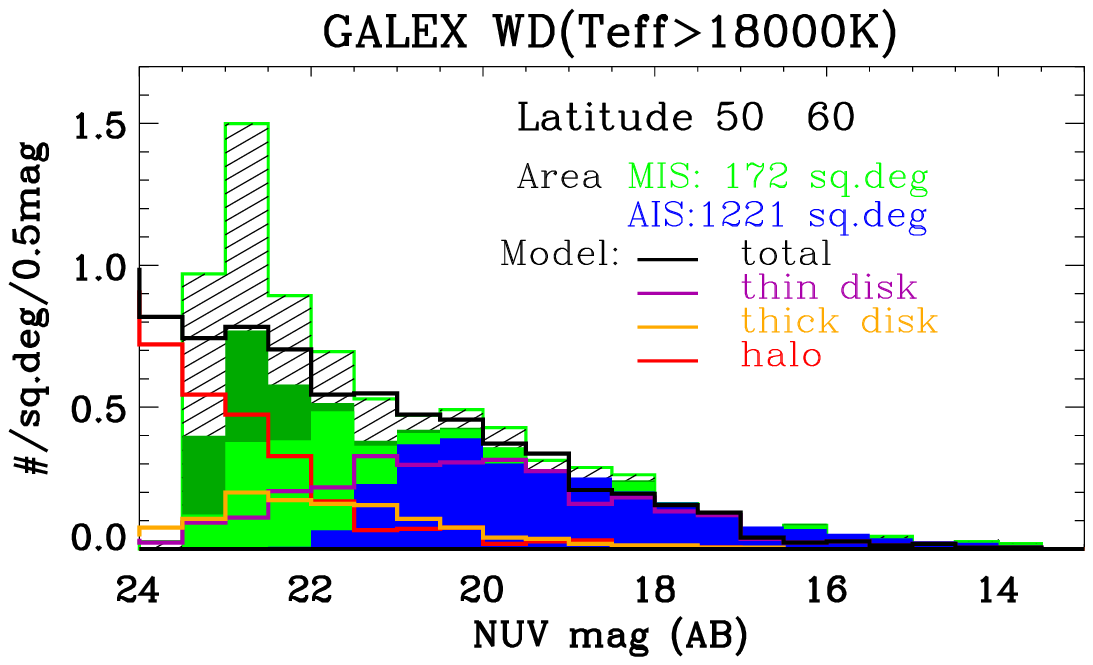}
\end{center}
\caption{\small Modeling WD counts: the effect of the IFMR. The density of hot star candidates 
at two  sample  latitudes (10$^{\circ}$-wide strips) 
are shown, with model predictions (halo, thin and thick disk components, and total)
The models in the top-left panel are  computed 
 with the TRILEGAL default IFMR  (MG07)  
and  the others with 
W2000 IFMR. The thick disk and halo counts  become significant
 at magnitudes fainter than $\sim$19, and $\sim$20, 
for these two IFMR respectively. 
Models with  W2000 IFMR 
match better the data down to magnitudes $\sim$20-21,  below which the AIS becomes incomplete. 
In the top and bottom-right  panels the model  counts are obtained selecting stars with \teff$>$18000~K  in the model calculations,
and in the bottom-left panel  selecting stars 
by color cut (FUV-NUV$<$-0.13) with TRILEGAL's transformation from \teff to magnitudes, from the same calculations.
The comparison illustrates the uncertainties introduced by the transformation of the isochrones into magnitudes
via model atmospheres.
The filled-green histograms are GALEX MIS hot ``single'' sources 
with SDSS match (light green), 
and including sources with no SDSS detection (dark green). 
  \label{f_strip_few} }
\end{figure*}

We use a new version of TRILEGAL, which has been modified for the
purposes of this  analysis in several ways; most of these modifications
will be detailed in a subsequent paper (Zabot et al., in
preparation). Post-AGB stars, planetary nebulae nuclei and white
dwarfs of types DA and DB have been included using the evolutionary
tracks from \citet{VW94} and \citet{Althaus_etal01,
Althaus_etal05}, together with the \citet{Koester08} synthetic
spectra. In the present version we have considered only the DAs, since
they are the dominant type among hot WDs \citep{HansenLiebert03}. The
fraction of DBs is known to increase significantly for WDs cooler than
12000~K, thanks to convection in the He layers \citep{Bergeron_etal97,
HansenLiebert03}. Such cool WDs, however, are not included in our
selection of sources.
The  TRILEGAL code uses the Koester (2008) grids to assign magnitudes
to the theoretical stars. In Figure \ref{f_models_FUV_NUV}  we show a  comparison 
with magnitudes calculated from TLUSTY models, and in Figure \ref{f_strip_few}
we compare stellar counts obatined by selecting model stars by \teff and by synthetic colors, 
in order to illustrate the 
sometimes neglected effect of  the model-atmosphere  uncertanties. 

Following our preliminary findings of discrepancies between predictions and hot star counts
based  on earlier data releases (Bianchi et al. 2009b), the TRILEGAL code 
has been modified   to  also allow the choice of  initial--final mass
relation (IMFR) independent from the prescription adopted for the
previous evolutionary phases. In this way, the mass distribution of the white dwarfs does not
need to follow the  constraints imposed by the previous TP-AGB tracks,
which come from Marigo \& Girardi (2007).

Model predictions for sample Galactic latitudes are shown in Figure \ref{f_strip_few},
 computed with  default TRILEGAL parameters for the MW geometry
and the default initial-final mass relation (IMFR) from  Marigo \& Girardi  (2007, hereafter MG07),
as well as the \cite{Weidemann2000} IFMR (hereafter W2000).  
We show the thin disk, thick disk, and halo stellar components, 
 as well as the total predicted counts. 
The IFMR from  MG07 
alrgely overpredicts faint star counts,
while   the W2000  IFMR produces an overall better 
match  with the observed counts, at all Galactic latitudes 
(see also Fig.  \ref{f_alllat}), and was therefore adopted in all our calculations that follow. 
Given the relevance of the IFMR in the context of stellar evolution, we explain in the next section how it affects the 
hot-star count predictions, which can be tested by our data. 

\begin{figure*}
\begin{center}
\includegraphics[width=70mm]{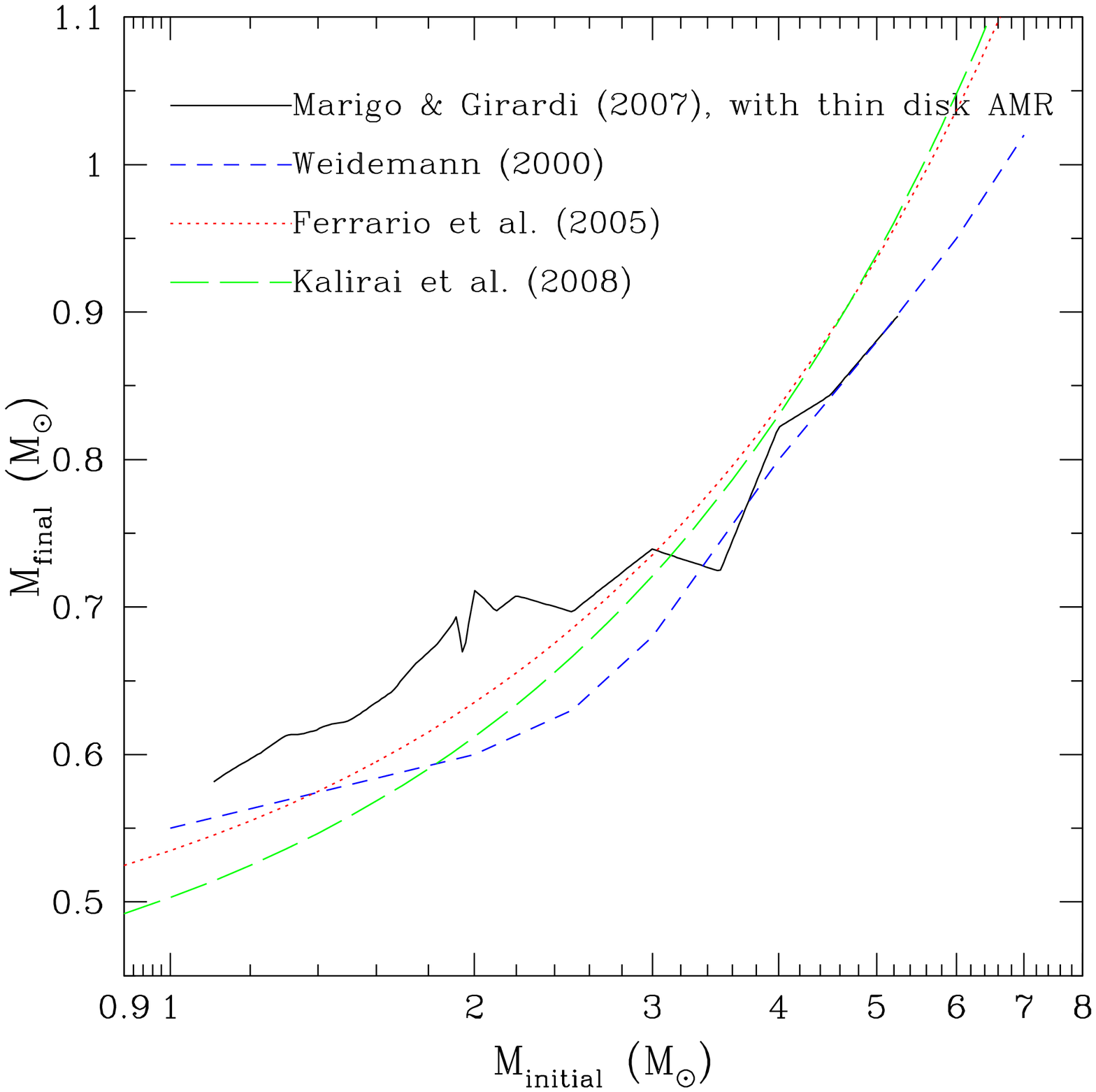}
\includegraphics[width=70mm]{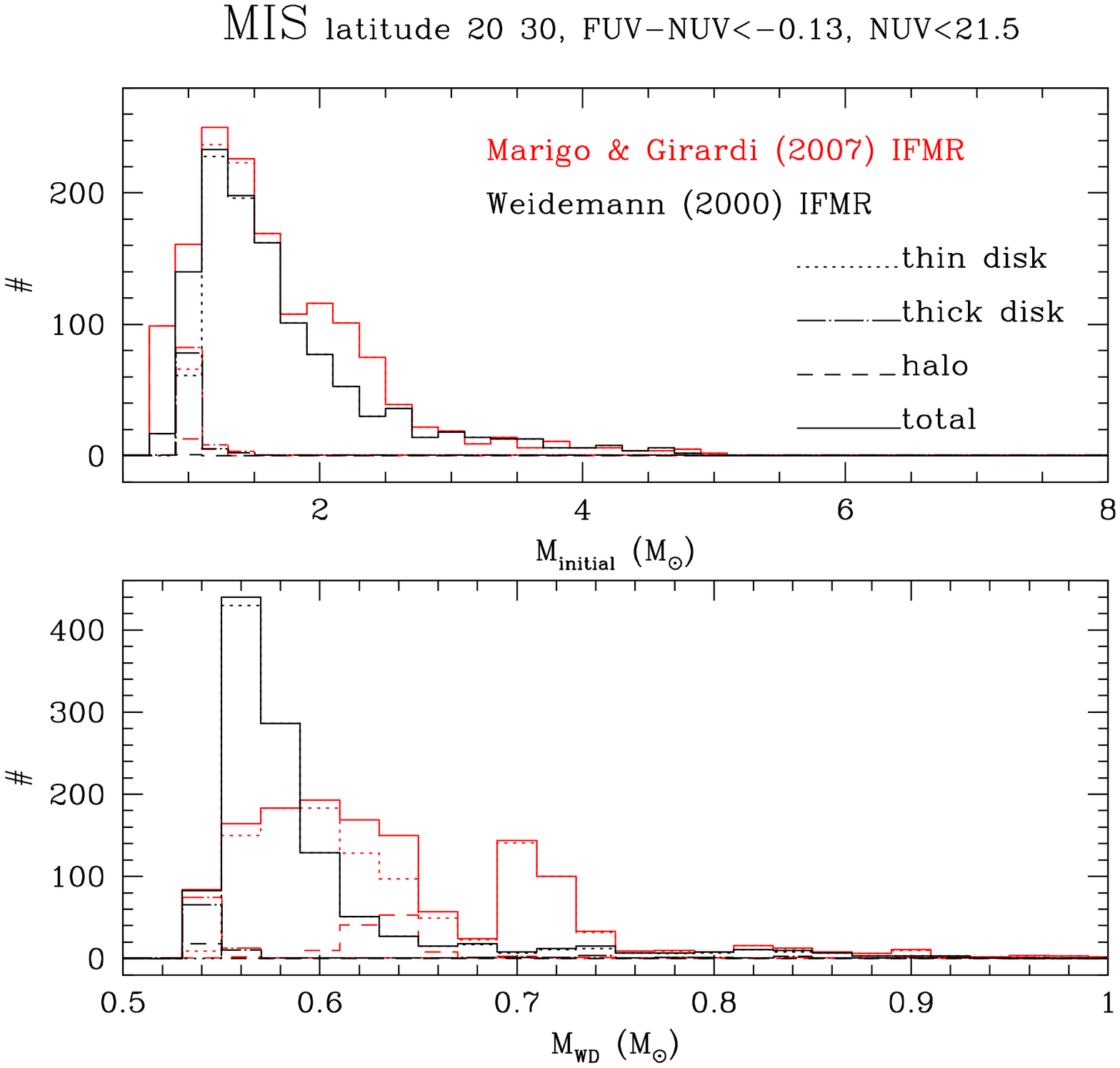}
\end{center}
\caption{\small {\bf LEFT:} Some of the IFMRs implemented in TRILEGAL, and tested in this 
work. The 
MG07 IFMR depends on metallicity, so that
different age--metallicity relations (AMR) will sample it in different
ways; the curve  shown here is for the \citet{RochaPinto_etal00} AMR,
which is used to model the MW thin disk in TRILEGAL. Other IFMRs are
the semi-empirical one from \citet{Weidemann2000}, and the purely
empirical ones by \citet{Ferrario_etal05} and \citet{Kalirai_etal08}. 
{\bf RIGHT:} The distribution of initial and final WD masses (top and 
bottom panels, respectively) derived from TRILEGAL models using the
\citet[][red lines]{MarigoGirardi07} and 
\citet[][black lines]{Weidemann2000} IFMR. The different lines 
illustrate the contributions from the different MW components, for one sample latitude, 
down to
NUV$=21.5$.  \label{f_ifmr} }
\end{figure*}

\begin{figure*} 
\begin{center}
\includegraphics[width=155mm]{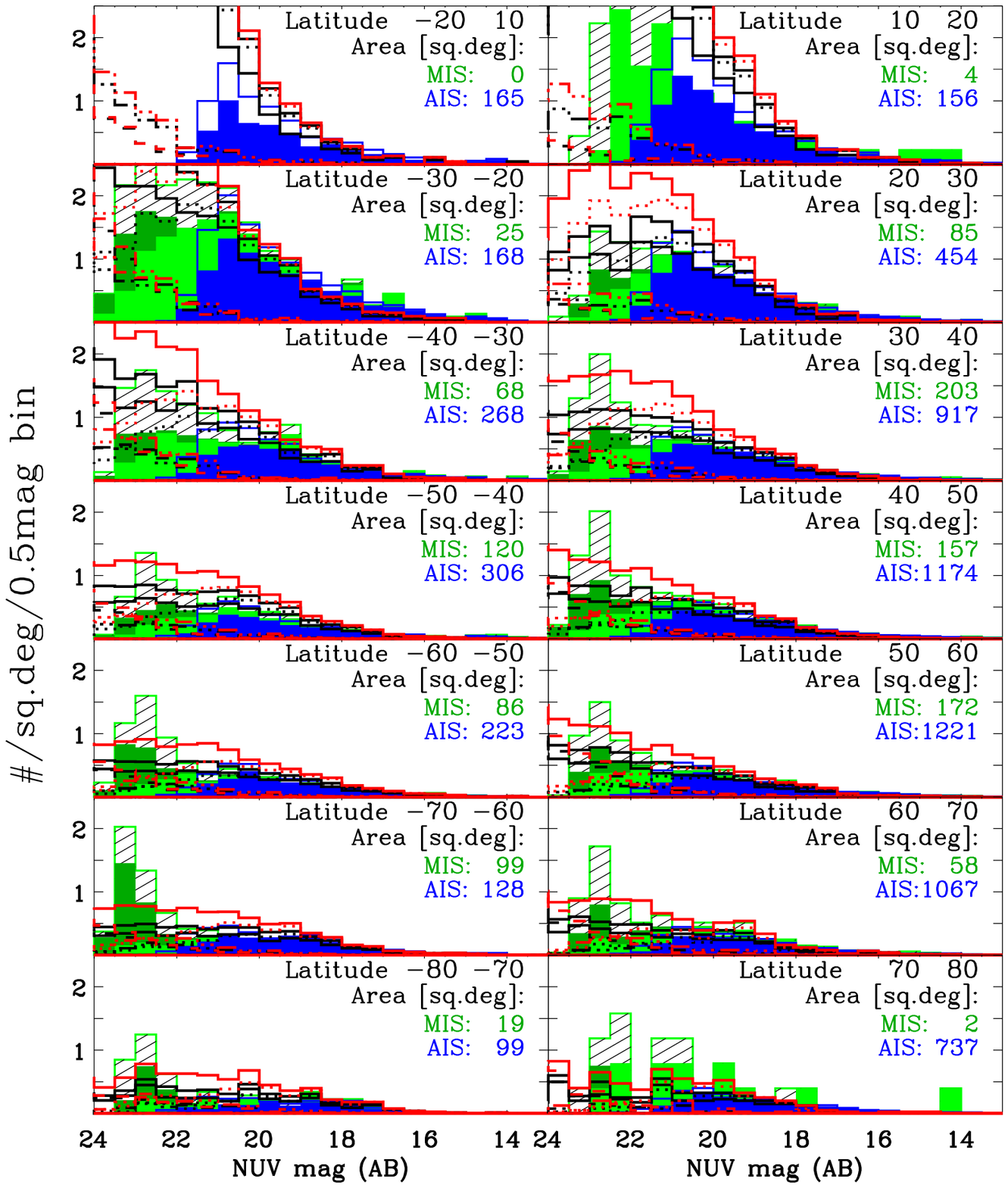}
\end{center}
\caption{\small Density of hot stars, separated in  
 10$^{\circ}$ ranges of Galactic latitude. The  area coverage for AIS and MIS, indicated on each panel, 
  varies significantly among latitudes,
and the statstics is better where the coverage is large.   
TRILEGAL model counts, computed with W2000 IFMR, 
are shown with  lines for thin disk (dotted), thick disk (dash-dotted)
and halo (dashed), as well as total (solid line). The black lines are model stars selected by \teff$>$18000K 
(upper line: total, lower line: single model stars); 
the red lines are stars selected from the same models by color cut (FUV-NUV$<$-0.13) as explained in Fig.8. 
 The match is  good overall, at bright magnitudes, but better at Northern high  latitudes and Southern low latitudes (except for $|b|$$<$20$^{\circ}$).
The dashed green-outlined histograms 
include both single and binary MIS hot-star candidates (the latter 
including also QSOs at faint magnitudes). 
The solid-color histograms are the ``single'' hot-star candidates (NUV-r$<$0.1).
The models were computed 
covering the centers of the GALEX MIS fields in each latitude strip, except for the lowest latitudes ($|$b$|$$<$20) where
the TRILEGAL models are computed following the distribution of the AIS fields, because MIS has little or no coverage. 
The large difference between North and South 20-30$^{\circ}$  latitudes, seen both in observations and predictions,
is due to the MIS-SDSS overlap including more directions towards the Galactic center in the South, and away from it in the North;
the longitude dependence is illustrated in Figure \ref{f_long}.   
  \label{f_alllat} }
\end{figure*}

\subsection{The Initial-Final Mass Relation}
\label{s_ifmr}
While the Milky Way geometry has been derived by previous studies of low mass stars, 
the IMFR is one of the least constrained factors in our understanding of stellar
evolution, in spite of its importance for  determining the yield of chemical elements,
and ultimately the Galaxy chemical evolution. The  reasons for this uncertainty include 
 insufficient statistics of  post-AGB stars. 
 The final WD mass is confined to a
very small range ($\sim$ 0.5-1.0\Msun)  compared to the $\sim$0.8-8\Msun range of their main sequence initial masses, 
and the post-AGB evolutionary time-scale varies steeply 
within the small WD mass range. The luminosity 
remains constant over a large range of \teff (up to $>$100,000K), and again depends on the remnant mass,
making it much harder to establish their distance and absolute luminosity than for  main 
sequence stars.

Therefore,  the mismatch between model predictions computed with TRILEGAL's  default parameters  and 
observed stellar counts (Figure \ref{f_strip_few}, top-left panel),
also previously noted by \cite{bia09fuseWD}, prompted us to explore  different
initial-final mass relations  in TRILEGAL. 
Although other factors, such as the WD birthrate, or the assumed MW extinction model,
or a different geometry 
may also affect model predictions, 
the IFMR is a most critical aspect for
modeling hot-WD counts, and was therefore explored in this first analysis. 
The geometry is better constrained by low-mass stars (e.g. Girardi et al. 2005), which are 
more numerous, and their counts based on optical-IR bands are less affected by extinction than a UV-based catalog. 
The right-hand panels
of Fig. \ref{f_strip_few} show that the 
W2000 IFMR produces
a better match of the model calculations with the observed
stellar counts than the MG07 IFMR, and this trend is seen at all Galactic latitudes.

These two IFMRs are illustrated in Fig.~\ref{f_ifmr}, together
with  additional IFMRs recently derived by  \citet{Ferrario_etal05}
and \citet{Kalirai_etal08} 
from extensive high-S/N spectroscopy of white dwarfs in Galactic open
clusters. Despite the uncertainties related to the age-dating of the
clusters, their field contamination, and the possible dependences on
metallicity, they are somewhat similar to the \citet{Weidemann2000}
semi-empirical IFMR.  The 
MG07 IFMR, instead, is
derived from theoretical evolutionary tracks and, as shown in the
figure, yields significantly higher final masses for all
$M_{\rm ini}<2\Msun$.

In order to understand why  the model-predicted hot-WD counts depend on the IFMR, 
 let us first consider the distribution of WD masses in
present-day surveys of the Solar neighborhood. Empirical
determinations have always concluded that the WD mass distribution
presents a strong peak at low masses,  with a maximum close to 0.55
or 0.65~\Msun~ and  a well-defined tail of more massive WDs, as well
as a faint low-mass tail believed to be either observational
errors or the result of binary evolution (e.g. 
\citet{Bergeron_etal92, Bragaglia_etal95, Madej_etal04,
Liebert_etal05, Kepler_etal07, Hu_etal07, Holberg_etal08}) .

The origin of this peaked mass distribution can be readily derived
from basic population synthesis theory \citep[see][]{Marigo01,
Ferrario_etal05}. Given a volume-limited sample containing stellar
populations of all ages between $T=0$ and $T=10$~Gyr, the distribution
of WD initial masses will be given by
\begin{equation}
N(M_{\rm ini}) \propto \phi_M(M_{\rm ini}) \,
	\psi[T-\tau_{\rm H}(M_{\rm ini})] \,
	\tau_{\rm WD}(M_{\rm ini}) \,,
\label{eq_mini}
\end{equation}
where $\phi_M$ is the IMF, $\psi[T-\tau_{\rm H}]$ is the star
formation rate at the time of stellar birth $T-\tau_{\rm H}$, and
$\tau_{\rm H}$ and $\tau_{\rm WD}$ are the main sequence and WD
lifetimes, respectively. The first 2 terms in the right-hand side of this
equation represent the production rate (or, alternatively, the ``death
rate'') of evolved stars with different masses, in number of stars per
unit time. Considering a \citet{Salpeter55} IMF, $\phi_M\propto M_{\rm
ini}^{-2.35}$, and a reasonably-constant star formation rate $\psi$
over the galaxy history, the equation above indicates a marked peak of
the production rate at the minimum initial mass for the formation of a
white dwarf, which is about 1~$M_\odot$ for a galaxy age of
$T=10$~Gyr. For higher initial masses, the production rate should fall
as $M_{\rm ini}^{-2.35}$. This behaviour will be shared by any galaxy
model with nearly-constant star formation, independently of its IFMR.

Assuming  a constant and monotonic IFMR, $M_{\rm WD}(M_{\rm
ini})$, the WD mass distribution is given by
\begin{equation}
N(M_{\rm WD}) \propto \left( \frac{\diff M_{\rm WD}}{\diff M_{\rm ini} 
	}\right)^{-1}\, 
	N(M_{\rm ini}) \,.
\label{eq_mwd}
\end{equation}
therefore, for almost-linear and linear IFMRs like
the \citet{Weidemann2000}, \citet{Ferrario_etal05}, and
\citet{Kalirai_etal08} ones, the WD production rate is still expected
to behave like a power-law peaked at the smallest masses. This
smallest mass corresponds to a WD mass of $\sim0.55$~\Msun\ (see 
Fig.~\ref{f_ifmr}).

\begin{figure}
\begin{center}
\includegraphics[width=80mm]{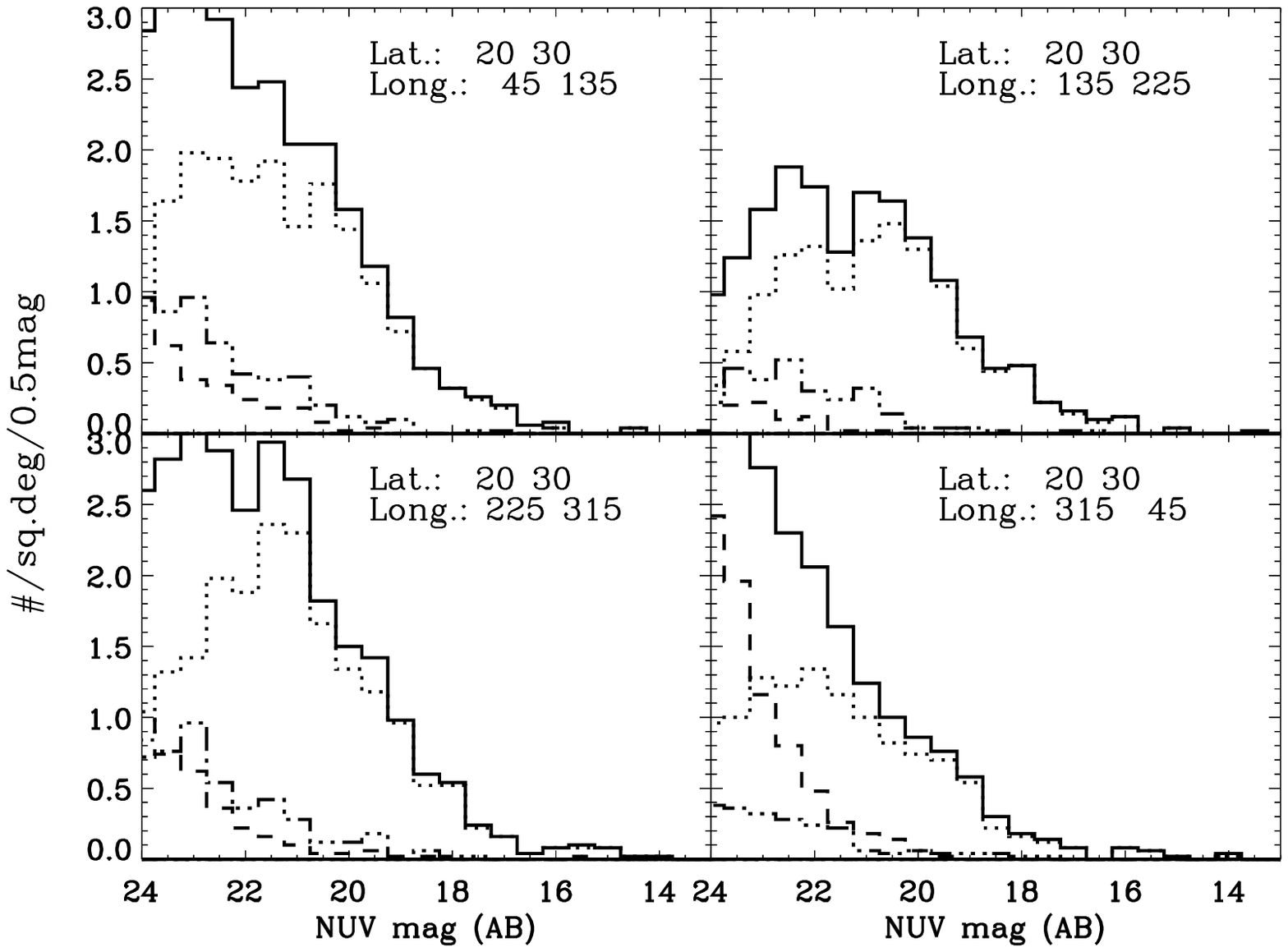}
\includegraphics[width=80mm]{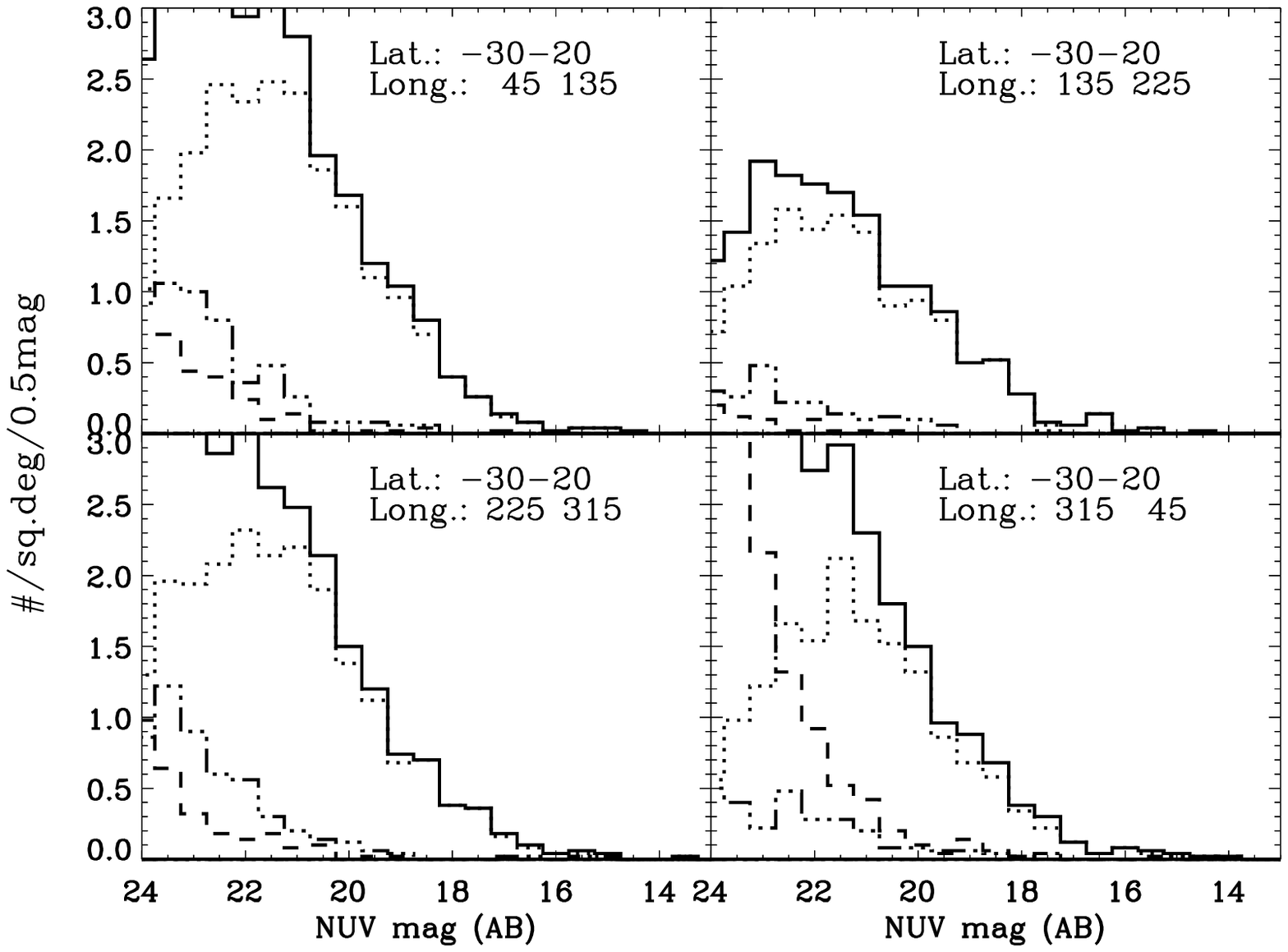}
\end{center}
\caption{\small 
Model-predicted counts at latitudes between 20-30$^{\circ}$ North, and South, for four symmetric longitude quadrants.
Model calculations were performed in 1$^{\circ}$ fields evenly distributed in longitude, and combined in 90$^{\circ}$ sections, to illustrate the dependence of
the model-predicted hot stars counts on longitude. The  strongest differences are seen between  center and anti-center directions, as
expected, and the  North-South asymmetry is most prominent in directions towards the Galactic center. 
Lines show thin disk (dotted), thick disk (dash-dotted), halo (dashed) and total counts (solid).
\label{f_long} }
\end{figure}

\begin{figure} 
\begin{center}
\includegraphics[width=80mm]{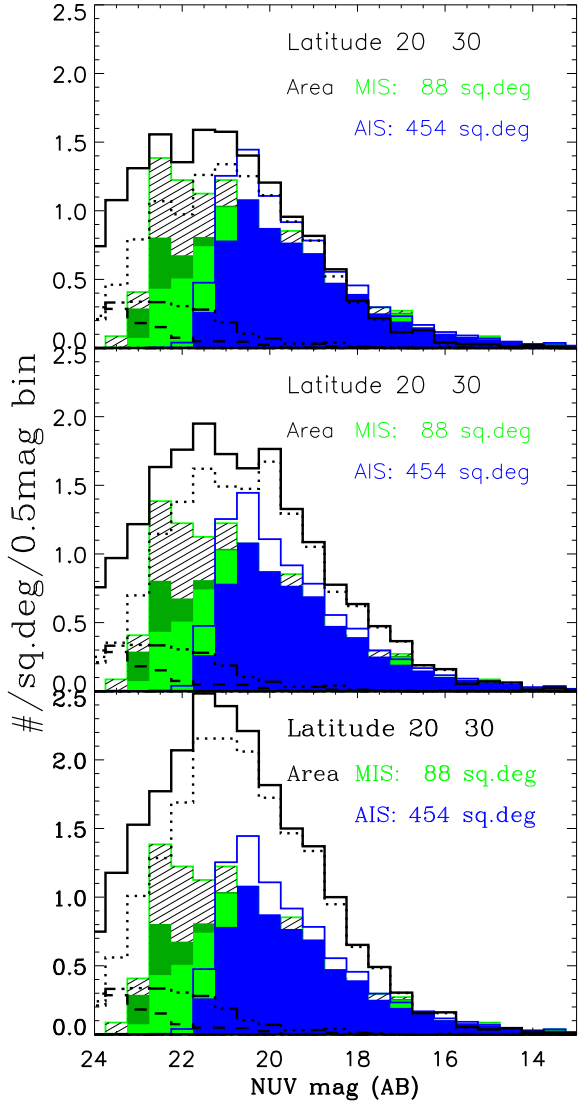}
\end{center}
\caption{\small Thin disk description. In view of the fact that our present data is most sensitive to the thin disk population,
we compare model results computed with  three different thin disk geometries:
in the top panel the TRILEGAL default geometry, 
in the middle panel an exponential (option 1), and in the bottom panel a z~sech (option 2) description (see text).
The top panel, with the default geometry description given in eq. 1 and 2, gives the closest results to the observed counts,
confirming previous findings by Girardi et al. (2005).  
This option was adopted in all our model calculations shown.
Colors and symbols as in Fig. \ref{f_alllat}.
\label{f_thind} }
\end{figure}

The WD lifetime $\tau_{\rm WD18k}$, defined as the total time after the AGB in which the
star will be hotter than 18,000K, is longer for WDs of higher mass, 
because, although they evolve much faster in the constant-luminosity phase, they reach higher \teff than  lower-mass WDs, 
and spend a longer time in their WD cooling track before
fading below the Teff=18000K limit.
Models using different IFMRs  simply
associate the final masses, hence also these  $\tau_{\rm WD18k}$ lifetimes,  to different initial masses, without
affecting the WD production rate. The end result in terms of mass
distribution $N(M_{\rm ini})$ (see equation 4) is that smaller
WD masses will weigh the 
 distribution towards 
shorter $\tau_{\rm WD}$, 
hence decreasing the numbers
of observed WDs with \teff$>$18000K. This effect explains the 
higher number of   hot WDs 
 predicted by the MG07 IFMR respect to the W2000 IFMR (Fig. \ref{f_strip_few}). 
 In fact, as can be inferred from  
Fig.~\ref{f_ifmr}, the 
W2000 IFMR  gives 
significantly lower WD final masses (hence shorter $\tau_{\rm WD18k}$) for the
 interval of initial masses between 0.6 and $\sim2$~\Msun. Because of the IFM being skewed 
towards lower masses, these objects represent numerically the majority within the whole mass range.

Figure \ref{f_ifmr} (right panel)  shows the distribution of initial and
final WD masses derived from TRILEGAL for  
a sample latitude, 
 using either the 
MG07 or W2000 IFMR. It shows that most of the hot WDs observed by
GALEX are expected to derive from low-mass stars, with initial masses
below $\sim2$~\Msun. Moreover, it indicates that the bulk of observed
WDs have low masses, typically $0.55\la M_{\rm WD}\la0.6$~\Msun\ in
the case of the W2000  
 IFMR. There is no direct
confirmation of this particular prediction, although it is in line
with WD mass determinations derived from spectroscopic surveys of more
limited samples \citep[e.g.][]{Bergeron_etal92, Bragaglia_etal95,
Madej_etal04, Liebert_etal05, Kepler_etal07, Hu_etal07,
Holberg_etal08}.

In the calculations with  MG07 IFMR 
the distribution of predicted WD masses is quite different from the
\citet{Weidemann2000} case, with a broad mass distribution extending from
0.55 to 0.65 \Msun, followed by a sort of gap, and a second peak of WD
masses located slightly above 0.7~\Msun. These features reflect the
positive slope of this IFMR up to $M_{\rm ini}\la2$~\Msun, and its
plateau at $M_{\rm WD}\simeq0.7$~\Msun\ for $2\la M_{\rm
ini}\la3.5$~\Msun\ (see Fig.~\ref{f_ifmr} -left). Another particularity
of this IFMR is the high  masses (about 0.64~\Msun) predicted for the
WDs belonging to the halo and thick disk. This prediction conflicts
with recent mass determinations of WDs in old star clusters, which
indicate values close to 0.53~\Msun\
\citep{Kalirai_etal08, Kalirai_etal09}.
A similar trend is seen at different latitudes.

Overall, an IFMR such as that of W2000 produces predicted counts closer to the observed numbers,
at bright magnitudes where our sample is complete, 
and  at the same time mass
distributions in agreement  with those derived from spectroscopy of
nearby WDs. These aspects underscore the difficulties of deriving the IFMR
from theoretical evolutionary tracks of AGB stars, even 
when these tracks are directly calibrated with observations of AGB
stars in the Local Group, as is the case of MG07.
This work, and future analyses planned on a wider sky area 
to explore concurrently the effects of the MW dust geometry, suggests that a deeper
investigation of AGB evolution is necessary, especially concerning
 the theoretical  prescriptions for mass loss, to reconcile constraints provided by 
WDs in our Galaxy and those provided by AGB stars in nearby galaxies. 

We finally  note that even models calculated with W2000 IFMR tend to predict more faint-star  counts than
our ``single'' WD sample (Fig.  \ref{f_alllat}). 
The mismatch is significant if we consider the red model lines (model stars selected by their synthetic FUV-NUV color),
which should be the most consistent  with the observed sample selection, to the extent that
the model atmospheres used by TRILEGAL are correct. 
Therefore, we explored also the case of 
a \citet{Kalirai_etal08} IFMR: the predicted counts are lower at faint magnitudes   (especially the
halo component fainter than NUV$\sim$22), as expected from Fig. \ref{f_ifmr}-left,  while still generally in line with   the bright
star counts. Because of the incompleteness of our sample at magnitudes  fainter than $\sim$21, 
and other factors that cannot be conclusively constrained with the current sample, 
we defer more quantitative  conclusions which rests on a  comparison of faint stars to 
a future work, and we adopt in this work  the W2000 IFMR, that matches well the bright star counts,
to explore other effects.

\subsection{Milky  Way geometry}
\label{s_geometry}
As explained above, the default parameters, in particular the MW
geometry, in the TRILEGAL code were defined from previous applications of this
code to analysis of stellar counts from several surveys. Particularly
important in this context were the 2MASS star counts, which are
sensitive to the low-mass stars ($<$0.8\Msun). These low-mass stars do
not overlap with the mass range of the hot-WD progenitors (about
 0.8-8\Msun), however they are very numerous and long-lived, and are
little affected by uncertainties in stellar evolutionary
models. Therefore we should expect the geometry to be well constrained
by previous works, 
except for the limited extent of the previous surveys.

The agreement between  our data  and  model predictions 
(with W2000 IFMR) is generally  good at bright magnitudes, where our sample is complete.
In more detail, we should also explore
the dependence of observed and predicted stellar counts
with longitude. The effect is more prominent towards the Galactic plane, 
therefore we examined latitudes between 20-30$^{\circ}$ North and South, 
and computed models with evenly spaced coverage along these strips.
The models predict, as expected, different counts towards the Galaxy center and anti-center,
shown in Fig. \ref{f_long}. The coverage of our present GALEX-SDSS matched catalog at low latitudes
is still too sparse for a conclusive comparison, which will be instead attempted 
by matching the AIS data with GSC2, and with the increased MIS 
coverage expected in the GALEX ``Extended Mission'' phase.

Finally, because our UV-optical matched samples  are complete to UV magnitudes brighter than
$\sim$20-21, therefore mostly sensitive to the thin disk population (Fig.s \ref{f_strip_few} and \ref{f_alllat}),
and because the models with default MW geometry slightly overpredict
the stellar counts at low (especially Northern) Galactic latitudes, we explored  different
geometries of the thin disk with  TRILEGAL.  In particular, in
Fig. \ref{f_thind} we show the effect of adopting more strongly peaked
density distributions at the Galactic Plane: either the simple
vertical exponential law, $\propto \rho^{z/h_z}$ (which has been
adopted in many MW models, starting from the classical
\citet{BahcallSoneira80, BahcallSoneira84}'s), or the simple
hyperbolic secant law, ${\rm sech}(2z/h_z)$.  The results confirm that
the default geometry adopted by \citet{Leo05}  
and in this work, i.e. the
squared hyperbolic secant law, provide the closest prediction to the
observed counts. The same distribution does also provide a good
match to star counts of ``normal stars'' in 2MASS and in the
local Hipparcos sample \citep{Leo05}. 
  The slight
overprediction and underprediction of hot star counts at certain
latitudes may also be due to extinction by interstellar dust, which may be
either patchy (as in fact it is) or have a slightly different
distribution than assumed. These remaining discrepancies and the effects of dust distribution may be addressed
with a wider MIS coverage in the future.

\subsection{The kinematic of Milky Way stars}
\label{s_kine}

\begin{figure*}
\begin{center}
\includegraphics[width=80mm]{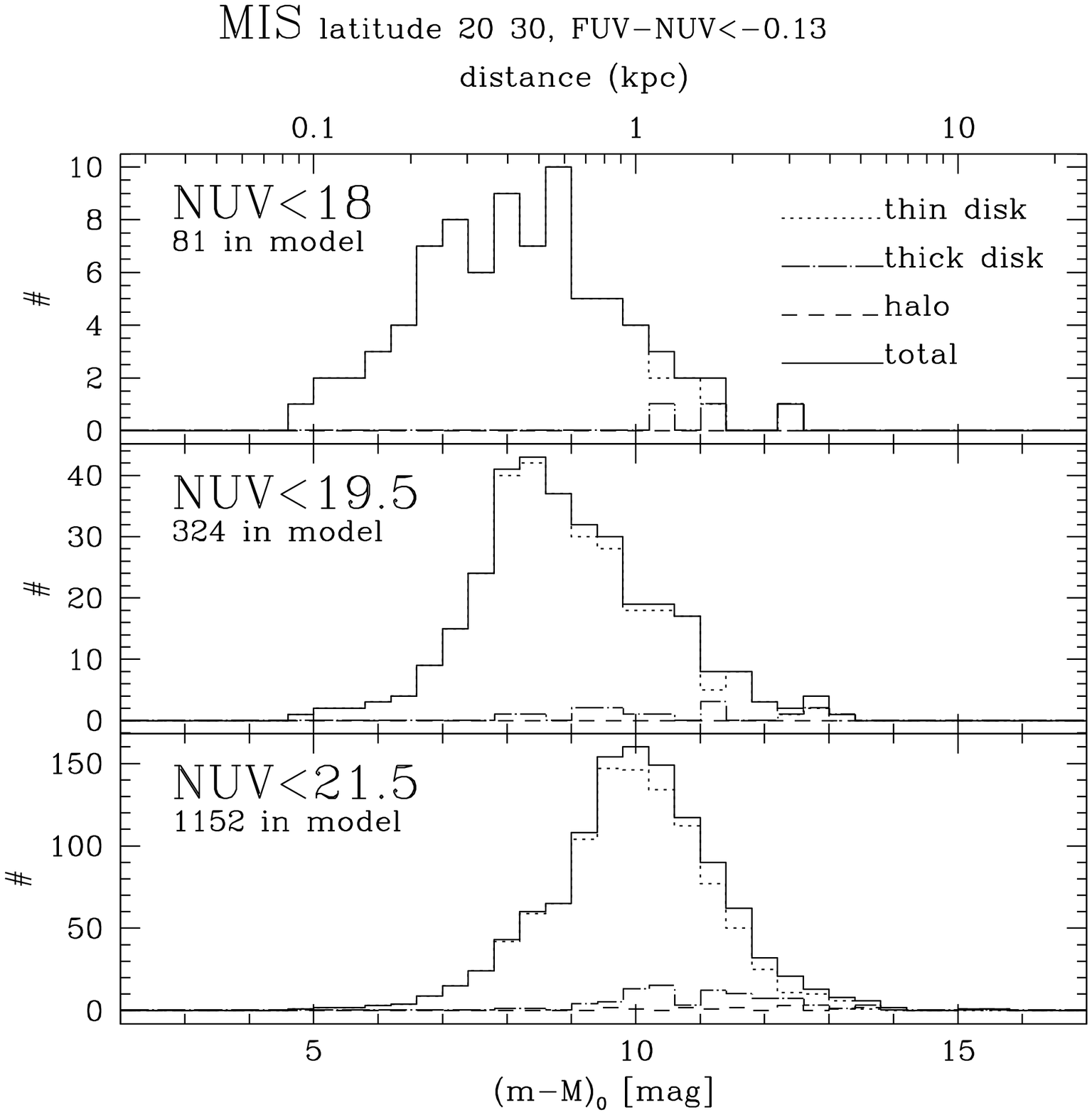}
\includegraphics[width=75mm]{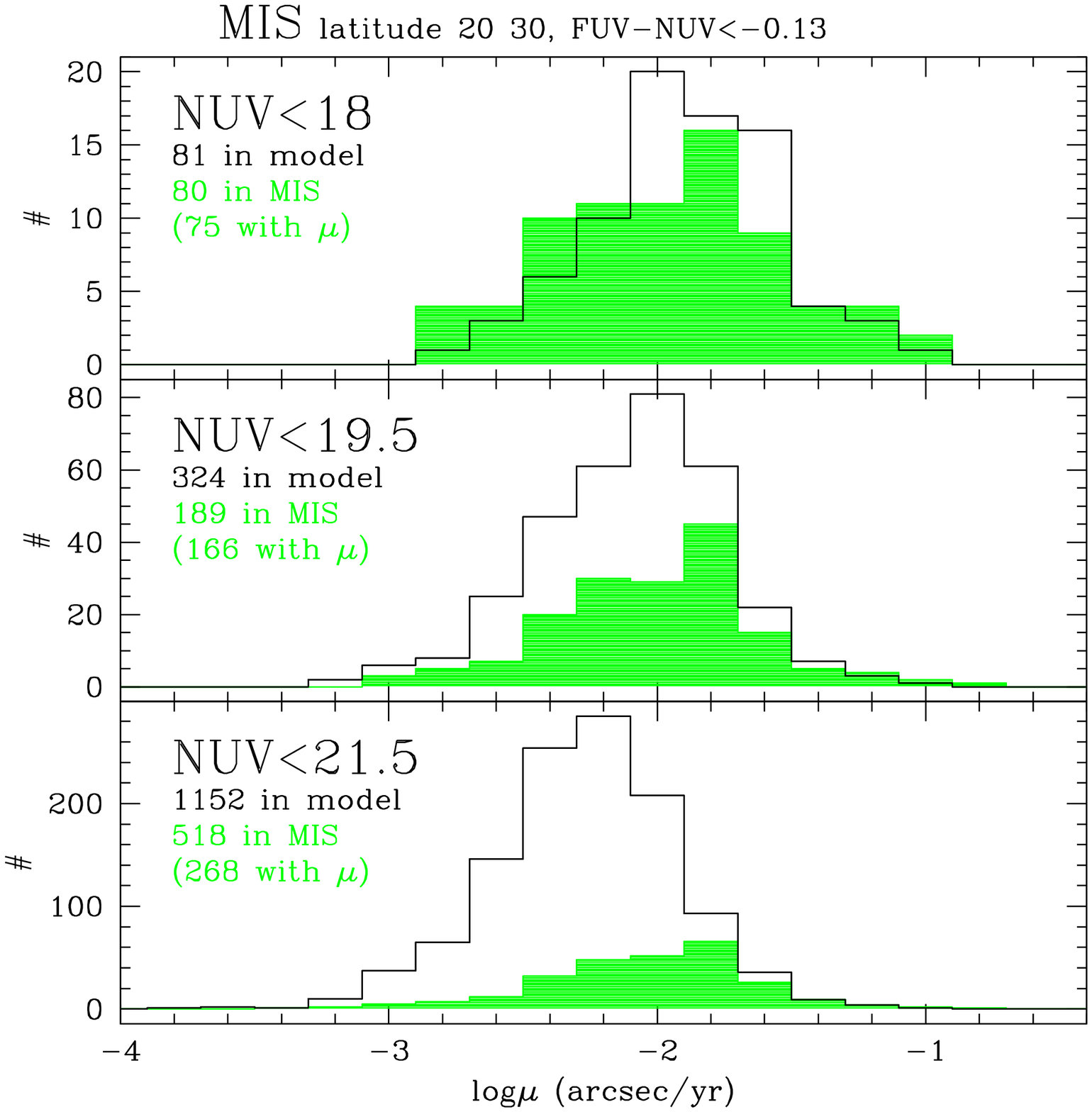}
\end{center}
\caption{\small Left: Distance distributions predicted by our TRILEGAL models
  for the hot stars in the 20-30N strip of MIS,
and for three different cuts in  NUV  magnitude. This plot indicates that the brightest (and complete) MIS sample of NUV$<$18mag is likely made of WDs
located in the thin disk at typical distances between 0.15 and 1 kpc (in this latitude range). The faintest samples 
could instead contain a significant fraction of stars at distances larger than 2 kpc,
comprising also a small fraction of thick disk and halo WDs. 
Right: distribution of proper motions (amplitudes):  observed
distribution  (green solid histograms) are compared to model predictions (continuous lines), computed for
the same area. 
 The numbers in each panel
indicate the NUV  magnitude cut, the number of predicted hot stars, of
observed ones in MIS, and of those which have proper motion information from USNO-B.
\label{f_pm} }
\end{figure*}

\begin{figure*} 
\begin{center}
\includegraphics[width=80mm]{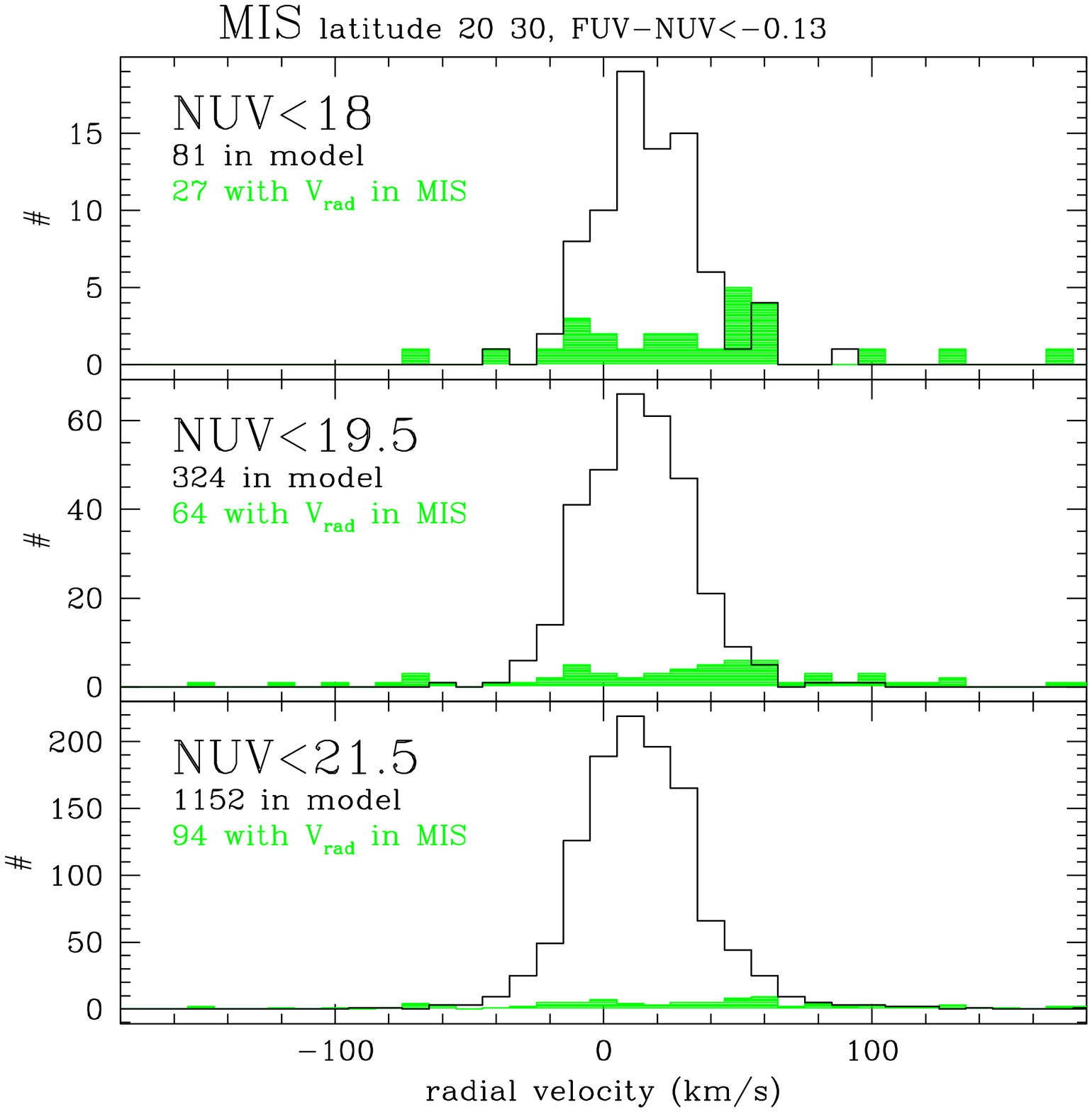}
\includegraphics[width=80mm]{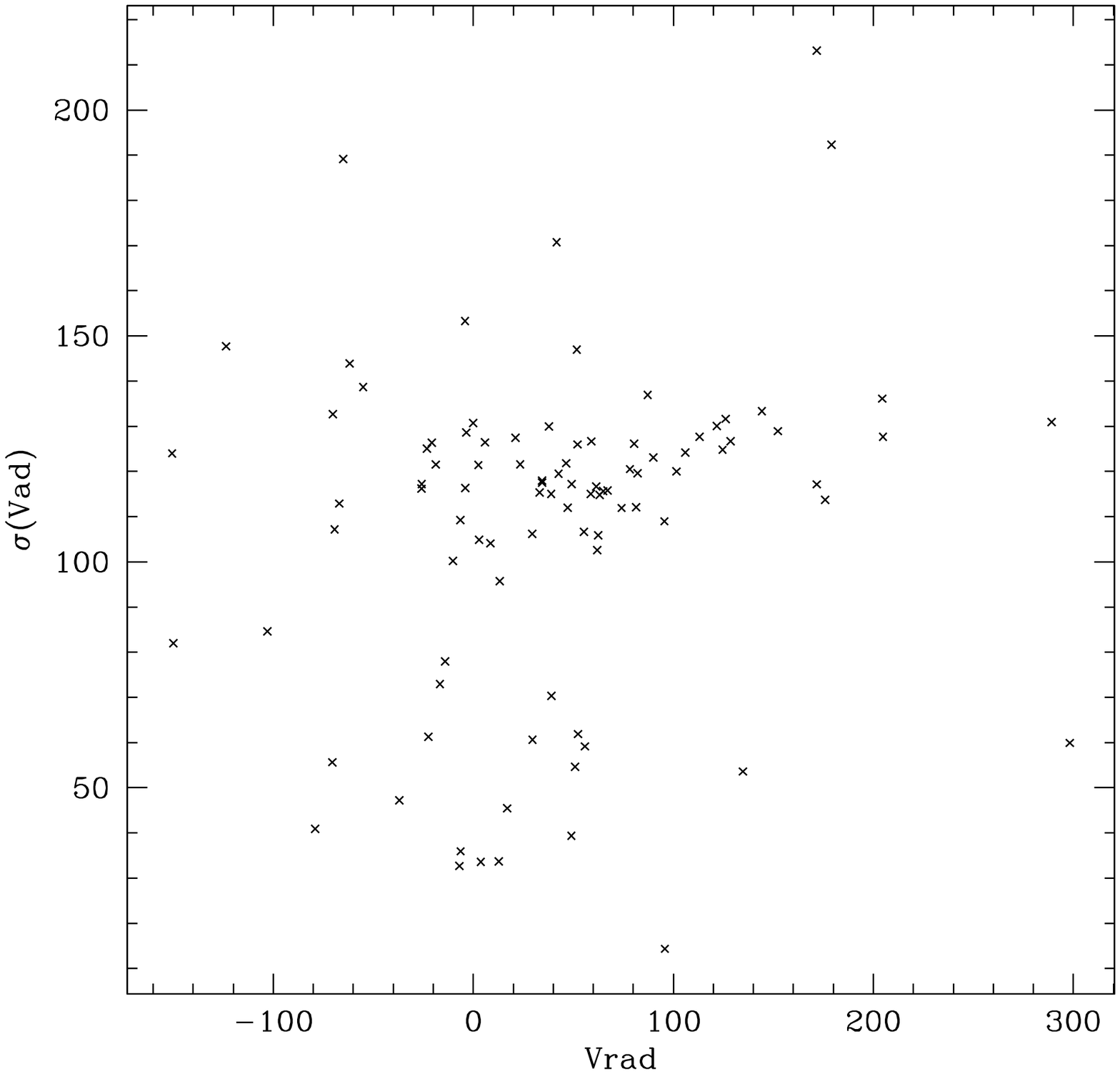}
\end{center}
\caption{\small Distribution of velocities (left) from the SDSS spectroscopic pipeline
for the subsample of MIS hot stars with spectra. 
 The comparison is
inconclusive because of the large errors (often larger than the values) in the
pipeline velocity measurements, shown in the right panel.   
\label{f_kine} }
\end{figure*}

The plots of stellar density versus Galactic latitude  presented in the previous
Sections clearly indicate that we are looking at a population of
objects for which the mean distance is larger than the mean scale
height on the thin disk. At typical distances of a few
hundred parsecs, there will be no parallax measurements for such
stars. The most direct distance information we can find for them
is related to their proper motions.

The table ``USNO'' in the SDSS database contains proper motions for SDSS objects made
after recalibrating USNO-B1.0  (\cite{Munn_etal04}) against SDSS astrometry.
The systematic errors are quoted to less than 1 mas/year. 
The  UCAC3 proper motion catalog is strongly
incomplete for our sample, owing  to its magnitude limit of
$f.mag\sim17$ (which roughly corresponds to $r\sim17$). In
fact, we checked and found that just a handful of our hot-WD candidates is
present in UCAC3.

Figure~\ref{f_pm} presents the histograms of proper
motions, $\mu$, for the hot stars in the 20-30N strip of MIS, and for
three different cuts in  NUV magnitude. The observed distributions
are compared with those predicted from the TRILEGAL models, computed for
the same area and for the W2000 IFMR. 
The model distributions are derived as follows.  
The  observed velocity ellipsoids 
$(\sigma_U,\sigma_V,\sigma_W)$ of each Galaxy component are taken form the literature. Each
simulated star is then given a random space velocity that follows the
Schwarzschild distribution for this ellipsoid. Space velocities are
then corrected for the solar motion (cf. Dehnen \& Binney 1998) and
projected on the sky using the transformations derived from Hipparcos. 
The final results are simulated proper motions and radial
velocities. We adopt an age-dependent velocity ellipsoid for the thin
disk, derived from the Geneva-Copenhagen Survey of Solar neighbourhood
(Holmberg et al. 2008). For the thick disk and halo, velocity
ellipsoids are taken from Layden et al. (1996). These
prescriptions provide a first-order description of the general
distributions of proper motions and radial velocities, but are not
expected to describe the details of the local velocity streams, nor
the effects caused by disk shear in the more distant stars. Moreover,
 we do not simulate measurement errors in the proper
motions. More details will be given elsewhere. 

The upper panel of  figure \ref{f_pm}-right compares the observed and model
distributions for a bright WD sample, with a cut at NUV$<$18mag. There is
quite good agreement both in the numbers and distributions of proper
motions. The agreement in the numbers derives from the fact that the
photometry is quite complete, and just 4\%  of this bright subsample does not have a proper motion listed in
USNO-B. This means that both  photometry and proper motion
information are very complete in this case, and quite similar to the
ones expected for a thin disk population.

The middle and bottom panels of  Fig.~\ref{f_pm}-right 
compare the observed and simulated $\mu$ distribution for fainter samples, with cuts at
NUV$<$19.5 and NUV$<$21.5mag, respectively. It shows that these samples
 become incomplete, first in the proper motions at 
NUV=19.5, and then in photometry at magnitudes NUV=21.5. Note for
instance the increasing fraction of stars without proper motion measurement in USNO-B,
which goes from 13 to 54\%  from NUV$<$19.5 to
NUV$<$21.5.

From this comparison  we conclude that our hot WDs do follow the
kinematics expected for the thin disk, even if the proper motion
information is severely incomplete for the faintest WD candidates.

Distance distributions predicted by the same models are shown in Fig. \ref{f_pm}-left. 
 This plot indicates that the brigthest (and complete) MIS sample of NUV$<$18mag is likely made of WDs
located in the thin disk at typical distances between 0.15 and 1 kpc in the latitude range 20-30$^{\circ}$  North. The fainter sample 
could instead contain a significant fraction of stars at distances larger than 2 kpc,
comprising also a small fraction of thick disk and halo WDs. Note however that this distribution of distances
is inferred from models, and we presently have no data to verify the distances of our sample stars, but for the
proper motions already depicted in Fig. \ref{f_pm} (right); future work will address identification of 
thick disk and halo objects with follow-up spectroscopy. 
The distance distribution is fairly similar at other latitudes. 

Finally, Fig.\ref{f_kine}  presents  the expected
distributions of radial velocities for the latitude 20-30N  sample,  from
the models, and the radial velocities derived from SDSS spectra,
available for a subsample of our hot star catalog (section \ref{s_purity}).
This is a biased subset
because the choice of SDSS spectroscopic targets is obviously unrelated to 
our selection of hot stellar objects. The worse effect however, 
making the comparison inconclusive, is the large errors in velocity determinations
from the SDSS pipeline (see Fig. \ref{f_kine}-right).

\section{Conclusions and Summary}
\label{s_conclusions}

From GALEX's data release GR5 we constructed catalogs of unique UV sources (i.e. eliminating repeated 
observations), 
and subcatalogs with matched SDSS optical photometry. We
extracted  catalogs of hot star candidates (FUV-NUV$<$-0.13),  mostly comprised of hot WDs. 
Over 38,000 such sources with photometric error $\leq$0.3mag,
 have  SDSS photometry ($\sim$74,000 with UV photometric error $\leq$0.5mag),
excluding UV sources with multiple optical matches,
that are between $\sim$8 to 30\% of the total.
The GALEX  surveys AIS and MIS
  cover different latitudes  at their respective depths ($\sim$21 and $\sim$23~AB mag), enabling a 
first quantitative analysis of the hot-WD stellar population with Milky Way models. 
Current descriptions of  MW stellar components (halo, thin and thick disk) 
are mostly based on stellar counts of low mass ($<$0.8\Msun) stars
(2MASS, shallow but all-sky, plus a few deep but area-limited surveys). 
We explored in this work different thin disk geometries and different IFMR. 
GALEX provides an unprecedented census of the evolved descendants of
 0.8-8\Msun stars, and the analysis 
of our current hot-WD sample  places some constraints on the initial-final mass relation (IFMR), 
one of the  crucial 
ingredients to understand the chemical enrichment of the ISM.

The brightest (and complete) sample 
 is likely made of hot WDs
located in the thin disk at typical distances between 0.15 and 1 kpc. The faintest samples 
could instead contain a significant fraction of stars at distances larger than 2 kpc,
comprising also hot WDs in the  thick disk and halo. 
This work is based on the analysis of matched GALEX-SDSS sources, and the SDSS magnitude  limit prevents
analysis of thick-disk and halo components. 
Hot WDs are however detectable at such larger distances
at MIS depth (see dark-green histograms in Fig.s \ref{f_strip_few} and \ref{f_alllat}, and section \ref{s_additions}),
and will be investigated in a future work with deeper optical data. 

 Model predictions of hot star counts  at different latitudes, computed with  currently accepted 
MW geometry, and assuming W2000 IFMR, match the data  quite well at intermediate 
latitudes and bright magnitudes, but are somewhat discrepant at low latitudes and faint magnitudes. 
There is a slight but systematic North-South asymmetry,  and MW models with canonical geometry for stars and dust
match better the data at Southern low latitudes, and at Northern high latitudes,  indicating that the dust distribution 
on the MW disk may also be better constrained by our data than was previously possible. 
The dust distribution  
will be investigated with the future more extended sky coverage from the ongoing GALEX mission. 
However, the observed asymmetry may also be due to the longitude dependence 
of the stellar populations, 
as our present matched-sources sample covers unevenly North and South Galactic latitudes (Fig. \ref{f_coverage}).
Our MIS matched-sources sample is incomplete at magnitudes fainter than about $\sim$21.5 due to the SDSS limits
for the hottest sources. 
At all latitudes, and at magnitudes where the samples are complete, MW model predictions of hot WD counts computed with 
 IFMRs favouring lower final masses 
match the GALEX hot WD counts significanly better 
than the IFMR currently postulated by  Marigo \& Girardi (2007) from  AGB stars data. 
This result contributes  an
important constraint to stellar evolution in the final phases which drive the yield of chemical elements.  

\subsection{The implications for stellar evolution} 

What are the implications of constraining the IFMR for the galaxy chemical evolution? 
The IFMR  contains the record of the
nucleosynthesis history during the TP-AGB phase,
as well as determines the contribution of AGB stars in
terms of dark remnants.
The IFMR, and the final yield of chemical elements, are 
significantly  affected by the occurrence (or not) of the so-called ``third dredge-up''.
At each  dredge-up episode the core mass is reduced by an
amount $\Delta M_{\rm dredge} = \lambda\, \Delta M_{\rm c,\,TP}$,
where $\lambda$ is the classical efficiency parameter, and
$\Delta M_{\rm c,\,TP}$ denotes the mass growth of the core
during the preceeding interpulse period.

In the extreme case where $\lambda
\sim 1$, no increment of the C-O core is predicted.
As a consequence,  for low-mass stars not experiencing the third
dredge-up ($\lambda=0,\, M\,< M_{\rm min}^{\rm dredge}$),
the remnant mass scales directly with
the duration of the TP-AGB lifetime, while for stars in which the third
dredge-up takes place ($\lambda>0,\,M\,\ge M_{\rm min}^{\rm dredge}$)
this simple proportionality is replaced by a more complex interplay
between core growth, dredge-up, and mass loss.
For such more massive stars
($M~\ge~M_{\rm min}^{\rm dredge}$, where $M_{\rm min}^{\rm dredge}
\approx~1.0-1.5~M_{\odot}$, depending on metallicity and model details)
the final mass and the chemical yields are related 
as we will briefly discuss below (see Marigo \& Girardi 2001 for more details).

For the generic element $k$, the corresponding AGB yield is defined as:
\begin{equation}
M_y^{\rm AGB}(k)=\int_{\tau_{\rm AGB}} [X_k(t)-X_k^0] \dot M(t)\, d t
\label{eq_yield}
\end{equation}
where $X_i^0$ is the initial abundance at the epoch of star formation,
$X_i(t)$ is the value at time $t$, and  $\dot M(t)$ is the current
mass-loss rate.

Let us first consider, in  the IFMRs in Fig. \ref{f_ifmr}, 
the part for $M_{\rm ini} < M_{\rm min}^{\rm dredge}$.
The chemical yields produced by these stars (mainly
$^{13}$C, $^{14}$N, and $^{4}$He) are
determined by the first dredge-up, and the cool-bottom process
(Boothroyd \& Sackmann 1999) or thermohaline mixing (Charbonnel \& Zahn
2007) operating during the RGB, as no subsequent 
dredge-up events take
place on the AGB according to the standard  stellar evolution theory.
For these low-mass stars the AGB yield, defined by
Eq.~(\ref{eq_yield}), simplifies as
$M^{\rm AGB}_y(k)~=~[X_k^{\rm RGB}-X_k^0] \Delta M_{\rm
AGB}^{\rm ej}$, where $X_k^{\rm RGB}$ is the surface abundance left
after the RGB phase, and  $\Delta M_{\rm AGB}^{\rm
ej}$ is the total amount of mass ejected on the AGB. In turn, the
expelled mass is related to the final mass through
$\Delta M_{\rm AGB}^{\rm ej} = M_{\rm AGB, 0}-M_{\rm final}$,
where  $M_{\rm AGB, 0}$ denotes the stellar mass at the onset of the AGB phase.
The final masses of these stars with
$M_{\rm ini} < M_{\rm min}^{\rm dredge}$ can be reduced by
simply invoking more efficient mass-loss  along the AGB.
Their chemical yields would increase correspondingly, 
since less envelope mass will be locked in the central core following the
outward displacement of the H-burning shell, so that
$\Delta M_{\rm AGB}^{\rm ej}$ will be larger.
For instance, for an AGB mass of  $M_{\rm AGB, 0}~=~1.2~M_{\odot}$,
a final mass $M_{\rm final} = 0.55~M_{\odot}$
versus  $M_{\rm final} = 0.60~M_{\odot}$
would imply an increase of the AGB chemical yield by
$(0.60-0.55)/(1.2-0.60)~\approx 10\%$, as well as a larger
total chemical yield, including the RGB wind contribution.

For $M_{\rm initial} \ge M_{\rm min}^{\rm dredge}$,
in addition to the first dredge-up (and possibly second dredge-up on
the Early-AGB),
the chemical yields include also the enrichment, mainly
in $^{4}$He and primary $^{12}$C, produced by the He-shell flashes and
brought up to the surface by the third dredge-up, and the enrichment
of $^{14}$N produced by the hot-bottom burning process occurring in the most
massive AGB stars ($M_{\rm initial} \ga 4.0-4.5~M_{\odot}$).
Again smaller final masses are reached when 
higher mass-loss rates occur, and/or  deeper dredge-up (i.e. larger
$\lambda$). These two factors may affect  the chemical yields in
opposite ways. On one hand, a high  mass loss limits the number of dredge-up
episodes, hence lowering the chemical yields; on the other hand, a deeper
dredge-up conveys larger amounts of carbon
and helium to the surface layer, which are  then ejected. In addition, 
mass loss and dredge-up efficiency 
are actually inter-related, 
a larger surface C abundance generally
favouring stronger dust-driven mass loss in C stars 
(e.g. Mattsson et al. 2008), while a significant reduction of the
mass of the envelope may decrease its penetration (i.e. lower $\lambda$) during
the third dredge-up (see, e.g. Karakas et al. 2002). In sum, a higher mass expelled 
does not directly imply that such material is  relatively more enriched of processed  elements. 

An IFMR similar to Weidemann (2000) weighs the final masses towards
lower values than e.g. the IFRM of MG07 in the range
$M_{\rm initial} \la 3.0 M_{\odot}$, while the two relations
essentially agree at higher $M_{\rm initial}$.
For $M_{\rm initial} \la M_{\rm min}^{\rm dredge}$,
forcing the theoretical IFMR to converge
on a W2000-type IFMR  by assuming a higher mass loss would
produce a modest increase of the chemical yields from these objects
(because more mass is ejected), compared to the MG07 IMFR case.
As for the  initial mass range
$M_{\rm min}^{\rm dredge}\le~M_{\rm initial}\le~3.0~M_{\odot}$, a higher mass loss
(necessary to reach a smaller final mass, as suggested by our WD counts) may
reduce the number of dredge-up episodes on the TP-AGB
and  this may lead to a lower yield of processed material.
In other words, constraining the IFMR translates into  constraining
 the total mass lost in the AGB phase.
The chemical composition of the ejecta from these stars, however,
is also critically sensitive to the
depth of the third dredge-up, which can be calibrated by
measurements of abundances in Planetary Nebulae
(Marigo et al 2003) and by a better estimate of the duration of
the AGB phase at different metallicity (Marigo et al. in prep.).

\subsection{Future Work} 
\label{s_future} 

With the current MIS coverage (1,103~square degrees overlapping with SDSS photometry in GR5),
 we compared with MW models stellar counts summed in 10$^{\circ}$  latitude strips, 
combining all longitudes in order to obtain  acceptable 
statistics. 
In order to disentangle geometry of the stellar populations, dust extinction, and stellar evolution in more detail,
we need to analyze also the  longitude dependence of stellar counts, as demonstrated by Fig.s \ref{f_alllat} and \ref{f_long}.
We will explore dust effects in the MW disk by matching the AIS with GSC2,
for a larger sky coverage (see \cite{biauvsky10}). The  density of hot stars (the rarest in nature) is very low and 
large area coverage is needed for good statistics. 

The MW halo and thick disk components, however,
 become significant at UV magnitudes fainter than about 21,
therefore require a wider MIS coverage 
(planned in the GALEX Extended Mission phase)
to be usefully constrained (see fig.s \ref{f_coverage}, \ref{f_strip_few} and \ref{f_alllat}),
and  deeper optical surveys. 

Finally, hot-WDs in binaries, again elusive at optical wavelengths, are 
uniquely and unambiguously revealed and characterized by the  GALEX UV sky survey, matched to optical surveys.
  These binaries 
sample  different types of stellar pairs (different \teff ranges) than those detected and
characterized by optical surveys, thanks to the UV photometry (e.g. Fig.6 of \cite{bia07b}).  
 Because our current photometric selection
includes contamination by extragalactic objects in the ``binary'' {\it locus},
and on the other hand the binaries do not significantly change the distribution of stellar counts, 
which was the subject of the present analysis,  
the binary candidates will be investigated in  a future work. 

\subsection{Final remarks. The online catalogs.} 
The present catalog of ``single'' hot star candidates has a high purity, as estimated
from the  serendipitous (but not unbiased) spectroscopic SDSS coverage.  While we expect most of the
hottest sources to be high gravity objects (CSPN, subdwarfs or WD), we did not apply any cut in 
gravity, therefore  main sequence  stars and supergiants hotter than $\sim$18,000K are also included
in our catalogs.  Their number is very small compared to the evolved stars, 
however their inclusion in the catalog makes it more generally useful. 

Finally, we remind future users of our catalogs that GALEX UV sources with more than one optical 
counterpart were excluded from the analysis of the  hot-star sample, 
(and should be excluded from any sample based on photometric color selection), but
 their fraction can be corrected statistically using Table \ref{t_stats}. 
They are included (flagged with ``rank''=1) in our online  catalogs. 
A few percent  of the matched  sources may be spurious matches (Fig. \ref{f_exptime2})
but these would mostly have 
random UV-optical colors. The hot-star candidates with NUV-{\it r}$>$0.1, 
includes also some  extragalactic objects, the relative fraction  is magnitude and latitude dependent.  

The source catalogs (GALEX unique sources, GALEX-SDSS matched sources, hot-stars samples)
are available in electronic form only, from the author's web site 
(http://dolomiti.pha.jhu.edu/uvsky ) where files description can be found,  
and will be also posted
on MAST (http://galex.stsci.edu and http://archive.stsci.edu/hlsp/). 
A statistical analysis and discussion of properties, useful to potential users of 
our (and similar) catalogs for understanding completeness and biases of 
any sample selection is given by \cite{biauvsky10}.

\section*{acknowledgments}
Data presented in this paper were obtained from the Multimission Archive at the Space Telescope Science Institute (MAST). STScI is operated by the Association of Universities for Research in Astronomy, Inc., under NASA contract NAS5-26555. Support for MAST for non-HST data is provided by the NASA Office of Space Science via grant NAG5-7584 and by other grants and contracts.

GALEX (Galaxy Evolution Explorer) is a NASA Small Explorer, launched in April 2003.
We gratefully acknowledge NASA's support for construction, operation,
and science analysis of the GALEX mission,
developed in cooperation with the Centre National d'Etudes Spatiales
of France and the Korean Ministry of 
Science and Technology. 
L.B. and J.H. acknowledge partial support from FUSE GI grant H901 (NASA NNX08AG97G).
A.Z. acknowledges financial support from CNPq-MCT/Brazil. 
L.G. acknowledges partial funding by from contract ASI-INAF I/016/07/0.
We are very greateful to A. Thakar for  discussions of very many issues   
regarding the SDSS database. 

\bibliographystyle{apj}

\begin{table}
\caption{Sky coverage for GR5  MIS and AIS, and GR5xDR7}
\label{t_area}
\begin{tabular}{rrrrr}
\hline
\multicolumn{1}{c}{Latitude}  &  \multicolumn{2}{c}{------- area [deg$^2$] -------} & 
 \multicolumn{2}{c}{E(B-V)$^{a}$}\\ 
\multicolumn{1}{c}{range}  & \multicolumn{1}{c}{GALEX GR5} & \multicolumn{1}{c}{GR5+DR7} & 
\multicolumn{1}{c}{mean} & \multicolumn{1}{c}{1$\sigma$ }  \\
\hline
MIS\\
\hline
$-90$  $-80$ & $   29.4$ & $    0.0$ &  n.a. & n.a. \\
$-80$  $-70$ & $   59.3$ & $   19.3$ &  $  0.04$ & $  0.01$ \\
$-70$  $-60$ & $  257.4$ & $   99.2$ &  $  0.03$ & $  0.01$ \\
$-60$  $-50$ & $  242.5$ & $  86.2 $ &  $  0.05$ & $  0.02$ \\
$-50$  $-40$ & $  150.5$ & $  120.4$ &  $  0.06$ & $  0.02$ \\
$-40$  $-30$ & $  109.0$ & $   68.04$ & $  0.07$ & $  0.03$ \\ 
$-30$  $-20$ & $   27.4$ & $   25.5$ &  $  0.08$ & $  0.02$ \\
$-20$  $-10$ & $    0.0$ & $    0.0$ &   n.a. & n.a. \\
$-10$  $  0$ & $    0.0$ & $    0.0$ & n.a. & n.a. \\
$  0$  $ 10$ & $    0.0$ & $    0.0$ &   n.a. & n.a. \\
$ 10$  $ 20$ & $    5.8$ & $    5.0$ &  $  0.04$ & $  0.01$ \\ 
$ 20$  $ 30$ & $   89.1$ & $   85.5$ & $  0.04$ & $  0.01$ \\ 
$ 30$  $ 40$ & $  210.3$ & $  203.2$ &  $  0.04$ & $  0.02$ \\ 
$ 40$  $ 50$ & $  161.3$ & $  157.1$ &  $  0.03$ & $  0.02$ \\ 
$ 50$  $ 60$ & $  175.7$ & $  173.0$ & $  0.04$ & $  0.01$ \\
$ 60$  $ 70$ & $   58.1$ & $   58.1$ &  $  0.03$ & $  0.01$ \\
$ 70$  $ 80$ & $    2.8$ & $    2.8$ &  $  0.02$ & $  0.00$ \\
$ 80$  $ 90$ & $    0.0$ & $    0.0$ &  n.a. & n.a. \\
  Total:  & $ 1578.6$ & $ 1103.0$ &  $  0.04$ & $  0.02$ \\ 
\hline
AIS\\
\hline
$-90$  $-80$ & $  249.3$ & $    0.0$ &  $  0.02$ & $  0.00$ \\
$-80$  $-70$ & $  714.0$ & $   99.6$ &  $ 0.03$ & $  0.01$ \\
$-70$  $-60$ & $ 1137.4$ & $  128.4$ &  $  0.03$ & $  0.01$ \\
$-60$  $-50$ & $ 1490.2$ & $  223.4$ & $  0.05$ & $  0.02$ \\
$-50$  $-40$ & $ 1752.0$ & $  306.8$ &  $  0.07$ & $  0.03$ \\
$-40$  $-30$ & $ 1910.2$ & $  268.8$ &  $  0.09$ & $  0.06$ \\
$-30$  $-20$ & $ 1616.3$ & $  168.7$ &  $  0.10$ & $  0.06$ \\
$-20$  $-10$ & $ 1015.9$ & $  100.9$ &  $  0.17$ & $  0.10$ \\
$-10$  $  0$ & $  211.3$ & $   37.6$ & $  0.41$ & $  0.24$ \\
$  0$  $ 10$ & $  349.5$ & $   27.4$ &  $  0.45$ & $  0.31$ \\
$ 10$  $ 20$ & $ 1487.0$ & $  156.3$ & $  0.12$ & $  0.14$ \\
$ 20$  $ 30$ & $ 2002.1$ & $  454.5$ &  $  0.06$ & $  0.05$ \\
$ 30$  $ 40$ & $ 2115.2$ & $  917.9$ &  $  0.05$ & $  0.02$ \\
$ 40$  $ 50$ & $ 1866.2$ & $ 1174.7$ &  $  0.03$ & $  0.02$ \\
$ 50$  $ 60$ & $ 1480.0$ & $ 1221.2$ &  $  0.03$ & $  0.01$ \\
$ 60$  $ 70$ & $ 1067.3$ & $ 1067.3$ &  $  0.02$ & $  0.01$ \\
$ 70$  $ 80$ & $  738.0$ & $  738.0$ &  $  0.02$ & $  0.01$ \\
$ 80$  $ 90$ & $  233.2$ & $  233.2$ &  $  0.02$ & $  0.01$ \\
  Total:   & $21434.8$ & $ 7324.5$ & $  0.05$ & $  0.07$ \\
\hline
\end{tabular}
\medskip{ 
(a)~\ebv values are computed from the \cite{Schlegel_etal98} maps for the centers of the GALEX fields, then averaged. They do not reflect 
an average Galactic trend, since they follow the specific distribution of fields shown in Fig. \ref{f_coverage}, which is rather non uniform at
low latitudes} 
\end{table}

\clearpage
\begin{landscape}
{\small
\begin{table} 
\caption{Statistical characteristics of the Catalogs and effects of error cuts}
\label{t_stats}
{\tiny 
\begin{tabular}{rrrrrrrrrrrrrrrrrrrrr}
\hline
\multicolumn{2}{c}{Latitude}  &                    
\multicolumn{2}{c}{\# GALEX}  & \multicolumn{2}{c}{\# GALEX sources}  &       
\multicolumn{6}{c}{--------------------- \# matched sources --------------------- }  &  
\multicolumn{9}{c}{------------------------ \# Pointlike matched sources ----------------------------}\\
\multicolumn{2}{c}{range} & 
\multicolumn{2}{c}{sources} & \multicolumn{2}{c}{FUV-NUV$<$-0.13} &
\multicolumn{2}{c}{err$_{NUV}$ $\leq$ 0.5} & 
\multicolumn{2}{c}{err$_{NUV,FUV}$ } &                      
\multicolumn{2}{c}{err$_{NUV,FUV}$ } &                      
\multicolumn{2}{c}{err$_{NUV}$ $\leq$0.5} &                 
\multicolumn{2}{c}{err$_{NUV,FUV}$} &                       
\multicolumn{2}{c}{err$_{NUV,FUV}$} &                       
\multicolumn{3}{c}{with  FUV-NUV$<$-0.13}\\ 
\multicolumn{2}{c}{} & 
\multicolumn{1}{c}{err$_{NUV}$} & \multicolumn{1}{c}{err$_{FUV,NUV}$} & \multicolumn{2}{c}{err$_{FUV,NUV}$} &
\multicolumn{1}{c}{rank0  } & \multicolumn{1}{c}{rank1  } & 
\multicolumn{2}{c}{ $\leq$ 0.5} &                      
\multicolumn{2}{c}{ $\leq$0.3} &                      
\multicolumn{2}{c}{} & 
\multicolumn{2}{c}{$\leq$0.5} & 
\multicolumn{2}{c}{$\leq$0.3} & 
\multicolumn{2}{c}{err$_{NUV,FUV}$}  & \multicolumn{1}{c}{NUV-r$<$0.1}\\
\multicolumn{2}{c}{} & 
\multicolumn{1}{c}{$\leq$0.5} & \multicolumn{1}{c}{$\leq$0.5} & \multicolumn{1}{c}{$\leq$0.5} & \multicolumn{1}{c}{$\leq$0.3} & 
\multicolumn{1}{c}{ }& \multicolumn{1}{c}{ } & 
\multicolumn{1}{c}{rank0  }& \multicolumn{1}{c}{rank1  } &               
\multicolumn{1}{c}{rank0  }& \multicolumn{1}{c}{rank1  } &               
\multicolumn{1}{c}{rank0  }& \multicolumn{1}{c}{rank1  } &    
\multicolumn{1}{c}{rank0  }& \multicolumn{1}{c}{rank1  } &    
\multicolumn{1}{c}{rank0  }& \multicolumn{1}{c}{rank1  } &    
\multicolumn{1}{c}{$\leq$0.5} & \multicolumn{1}{c}{$\leq$0.3} &  \multicolumn{1}{c}{$\leq$0.3}\\ 
\hline
MIS \\
\hline
-90 & -85 & 	 66929 & 16822 & 	    2444 &     622 &       0 &       0 &       0 &       0 &       0 &       0 &       0 &       0 &       0 &       0 &       0 &       0 &       0 &       0 &       0\\
 -85 & -80 & 	 177553 & 49043 & 	    6304 &    1540 &       0 &       0 &       0 &       0 &       0 &       0 &       0 &       0 &       0 &       0 &       0 &       0 &       0 &       0 &       0\\
 -80 & -75 & 	 112394 & 32092 & 	    4021 &     971 &       0 &       0 &       0 &       0 &       0 &       0 &       0 &       0 &       0 &       0 &       0 &       0 &       0 &       0 &       0\\
 -75 & -70 & 	 394332 & 109647 & 	   12984 &    3070 &   87456 &    2708 &   33147 &    1300 &   15510 &     794 &   17782 &     638 &    2584 &     211 &    1158 &     127 &     321 &      91 &      42\\
 -70 & -65 & 	 798670 & 223254 & 	   25020 &    6151 &  156597 &    3994 &   63005 &    2099 &   34584 &    1450 &   33311 &    1021 &    5554 &     345 &    2759 &     224 &     671 &     201 &      78\\
 -65 & -60 & 	 1409943 & 400165 & 	   44106 &   11391 &  360403 &   10991 &  149854 &    5826 &   87609 &    4079 &   71801 &    2433 &   12151 &     866 &    6395 &     565 &    1386 &     426 &     155\\
 -60 & -55 & 	 1244707 & 328909 & 	   40804 &   10773 &  228705 &    7106 &   86721 &    3448 &   49148 &    2388 &   49844 &    1659 &    7401 &     501 &    4061 &     334 &     931 &     296 &     125\\
 -55 & -50 & 	 632803 & 143121 & 	   21932 &    6162 &  164257 &    5294 &   49621 &    2093 &   25136 &    1291 &   41518 &    1388 &    4477 &     308 &    2374 &     204 &     711 &     257 &     112\\
 -50 & -45 & 	 609861 & 137541 & 	   21637 &    5756 &  304497 &   12522 &   92372 &    5102 &   46024 &    3217 &   77719 &    3427 &    8067 &     812 &    4342 &     506 &    1230 &     472 &     244\\
 -45 & -40 & 	 428767 & 89077 & 	   13169 &    3620 &  207041 &   10578 &   57421 &    3948 &   29654 &    2414 &   61074 &    3127 &    5178 &     651 &    2869 &     413 &     745 &     311 &     157\\
 -40 & -35 & 	 538838 & 103347 & 	   15242 &    4283 &  211133 &    9835 &   50373 &    3174 &   27961 &    1978 &   83015 &    3652 &    6030 &     580 &    3634 &     349 &     843 &     396 &     228\\
 -35 & -30 & 	 323258 & 59584 & 	    8370 &    2715 &  124660 &    8736 &   27139 &    2794 &   16141 &    1929 &   58244 &    3735 &    3658 &     607 &    2374 &     419 &     558 &     303 &     174\\
 -30 & -25 & 	 214275 & 31837 & 	    5398 &    1749 &  133209 &    8714 &   23127 &    2116 &   11900 &    1213 &   75788 &    4697 &    4076 &     529 &    2404 &     301 &     656 &     365 &     240\\
 -25 & -20 & 	 48433 & 6190 & 	    1035 &     357 &   34035 &    2638 &    4825 &     494 &    2441 &     272 &   22202 &    1652 &    1109 &     149 &     646 &      94 &     182 &      93 &      65\\
 15 & 20 & 	 43887 & 8626 & 	     997 &     304 &   26939 &    1383 &    6195 &     433 &    3488 &     279 &   14452 &     716 &    1297 &     120 &     812 &      81 &     132 &      77 &      51\\
 20 & 25 & 	 226867 & 47323 & 	    6099 &    1672 &  144508 &    7255 &   35983 &    2484 &   18325 &    1450 &   64767 &    3040 &    5810 &     575 &    3372 &     358 &     635 &     308 &     218\\
 25 & 30 & 	 403868 & 91523 & 	   10958 &    2782 &  281365 &   12461 &   79869 &    4695 &   39329 &    2745 &  101054 &    4591 &   10017 &     974 &    5594 &     565 &    1075 &     520 &     346\\
 30 & 35 & 	 711824 & 152925 & 	   21310 &    5682 &  474446 &   18387 &  132915 &    6920 &   66709 &    4116 &  155923 &    6218 &   14995 &    1263 &    8182 &     727 &    2079 &     862 &     488\\
 35 & 40 & 	 937945 & 203696 & 	   27963 &    7589 &  621803 &   24188 &  173805 &    9065 &   91210 &    5604 &  200400 &    8114 &   18511 &    1602 &   10080 &     971 &    2720 &    1094 &     601\\
 40 & 45 & 	 641846 & 152205 & 	   21044 &    5661 &  430026 &   15963 &  133123 &    6575 &   68076 &    4018 &  119976 &    4803 &   12413 &    1116 &    6306 &     654 &    1892 &     744 &     413\\
 45 & 50 & 	 707063 & 162317 & 	   21478 &    5659 &  460636 &   15730 &  139985 &    6496 &   74669 &    4238 &  127343 &    4598 &   12844 &    1041 &    6839 &     658 &    1885 &     639 &     325\\
 50 & 55 & 	 731762 & 170967 & 	   20334 &    5143 &  492531 &   19798 &  149059 &    7947 &   80742 &    4944 &  135058 &    5360 &   13351 &    1257 &    7468 &     769 &    1700 &     641 &     371\\
 55 & 60 & 	 690921 & 170036 & 	   20115 &    4864 &  461420 &   18063 &  146692 &    7491 &   76642 &    4663 &  119973 &    4788 &   12470 &    1150 &    6521 &     704 &    1657 &     528 &     293\\
 60 & 65 & 	 377644 & 97916 & 	   11366 &    2899 &  255057 &    9001 &   86150 &    4034 &   45134 &    2581 &   62828 &    2388 &    7169 &     649 &    3695 &     415 &     912 &     309 &     142\\
 65 & 70 & 	 99569 & 25651 & 	    3073 &     716 &   68009 &    2407 &   22683 &    1058 &   10337 &     609 &   13951 &     546 &    1857 &     151 &     832 &      83 &     238 &      75 &      44\\
 70 & 75 & 	 23953 & 6405 & 	     678 &     136 &   16778 &     537 &    5757 &     238 &    2588 &     137 &    3126 &     137 &     446 &      46 &     198 &      26 &      62 &      20 &      12\\
-90 & 90 & 	 12597912 & 3020219 &	  387881 &  102267 & 5745511 &  228289 & 1749821 &   89830 &  923357 &   56409 & 1711149 &   72728 &  171465 &   15503 &   92915 &    9547 &   23221 &    9028 &    4924\\
\hline
AIS\\
\hline
-85 & -90 & 186959 & 29800 & 	   6088 &     1187 &     2118 &     66&    488&     22&    137&      9&    686&     24&     60&      2&     21&      1&      5&      0&      0\\
-80 & -85 & 639403 & 107385 & 	  20863 &     4235 &    15996 &    448&   3458&    144&   1075&     65&   4635&    152&    399&     33&    177&     17&     55&     28&     24\\
-75 & -80 & 946695 & 157960 & 	  31268 &     6157 &    42060 &   1425&   9492&    431&   2998&    179&  11306&    416&    922&     79&    448&     32&    134&     60&     41\\
-70 & -75 & 1379695 & 220313 & 	  43504 &     8513 &   102750 &   4001&  23114&   1254&   7740&    600&  28113&   1238&   2273&    223&   1058&    116&    309&    138&     97\\
-65 & -70 & 1779025 & 267234 & 	  53565 &    10329 &   104406 &   3670&  24373&   1188&   8059&    514&  28907&   1114&   2464&    221&   1070&    102&    350&    141&     99\\
-60 & -65 & 1823298 & 273723 & 	  54203 &    10650 &   151592 &   6068&  34985&   1981&  11556&    944&  44240&   1714&   3466&    342&   1584&    181&    489&    233&    176\\
-55 & -60 & 2202425 & 309450 & 	  62380 &    12610 &   149424 &   6616&  29732&   1895&   9752&    819&  50822&   2148&   3580&    355&   1700&    168&    537&    294&    214\\
-50 & -55 & 2221574 & 260072 & 	  55813 &    11764 &   137391 &   6620&  21591&   1640&   7251&    732&  56334&   2497&   3270&    348&   1589&    154&    553&    331&    234\\
-45 & -50 & 2526007 & 271544 & 	  59711 &    12805 &   195821 &  10880&  29085&   2665&  10396&   1253&  81131&   4309&   4433&    562&   2221&    277&    801&    473&    338\\
-40 & -45 & 2644017 & 259939 & 	  59142 &    14759 &   157644 &  12248&  20715&   3038&   7274&   1400&  79249&   5390&   3708&    680&   1808&    341&    758&    441&    302\\
-35 & -40 & 2824496 & 243813 & 	  55332 &    13542 &   243682 &  15446&  28209&   3371&   9899&   1562& 133778&   7605&   5717&    831&   2900&    417&   1208&    725&    501\\
-30 & -35 & 3022905 & 251126 & 	  53008 &    14535 &   171132 &  14681&  14919&   2977&   5579&   1395& 112514&   7969&   4148&    857&   2175&    454&    938&    570&    394\\
-25 & -30 & 2725467 & 171367 & 	  39306 &    12449 &   219113 &  20491&  15759&   3593&   6299&   1665& 159944&  12263&   5423&   1154&   2978&    571&   1330&    829&    611\\
-20 & -25 & 2565319 & 121358 & 	  29997 &    11039 &   158390 &  19012&   7832&   2255&   3563&   1116& 133521&  13447&   4060&    922&   2233&    504&   1037&    650&    476\\
-15 & -20 & 2294738 & 90915 & 	  21775 &     9748 &   104999 &  13902&   4288&   1362&   2245&    739&  94135&  10745&   2685&    665&   1536&    384&    660&    406&    291\\
-10 & -15 & 1474126 & 54091 & 	  11682 &     6026 &   107441 &  14988&   4086&   1142&   2582&    762& 102078&  12821&   3235&    736&   2048&    511&    607&    350&    231\\
-5 & -10 & 468735 & 20256 & 	   3196 &     1719 &    15893 &   2690&    834&    224&    547&    147&  15334&   2358&    707&    158&    441&    102&     77&     39&     22\\
0 & -5 & 81792 & 5544 & 	    338 &      126 &        0 &      0&      0&      0&      0&      0&      0&      0&      0&      0&      0&      0&      0&      0&      0\\
5 & 0 & 112359 & 9822 & 	    673 &      324 &     4730 &    816&    364&    140&    263&    120&   4577&    774&    340&    133&    241&    113&     37&     18&      2\\
10 & 5 & 896414 & 41563 & 	   5055 &     2532 &    67009 &  10668&   4191&   1170&   2914&    871&  64784&   9577&   3683&    921&   2520&    682&    285&    169&     86\\
15 & 10 & 2209741 & 93883 & 	  17027 &     7825 &   133415 &  20698&   6792&   1979&   4160&   1267& 124034&  17107&   5085&   1207&   3230&    806&    758&    458&    301\\
20 & 15 & 2785094 & 137819 & 	  28011 &    10512 &   248149 &  22946&  22199&   3385&   9760&   1708& 192683&  16088&   8372&   1376&   4610&    747&   1398&    892&    688\\
25 & 20 & 3078141 & 180485 & 	  38516 &    11924 &   482621 &  34646&  50125&   5965&  19504&   2777& 328234&  21367&  14284&   2013&   7510&   1020&   2539&   1604&   1230\\
30 & 25 & 3074765 & 199400 & 	  45162 &    11949 &   565885 &  34039&  59037&   6033&  20866&   2622& 336893&  18763&  14374&   1854&   7217&    885&   2777&   1683&   1308\\
35 & 30 & 3068226 & 221555 & 	  50831 &    11949 &   770294 &  40366&  91294&   7455&  30887&   3131& 420587&  20138&  18413&   1879&   9082&    840&   3615&   2135&   1714\\
40 & 35 & 2827184 & 215451 & 	  50937 &    10501 &   911569 &  43682& 109497&   8095&  35498&   3333& 459207&  20279&  19905&   1846&   9697&    822&   3899&   2278&   1774\\
45 & 40 & 2867019 & 241576 & 	  57082 &    10742 &  1149694 &  53336& 144002&  10172&  45305&   4158& 490863&  21484&  21555&   2157&  10284&    980&   4171&   2345&   1858\\
50 & 45 & 2589087 & 237387 & 	  55351 &     9692 &  1187271 &  53682& 158334&  10737&  47012&   4333& 455468&  19957&  21918&   2243&  10131&   1013&   4017&   2139&   1679\\
55 & 50 & 2183650 & 219294 & 	  49511 &     8631 &  1135384 &  51155& 165181&  11031&  48979&   4336& 414632&  18158&  21966&   2136&   9994&    973&   3845&   1979&   1556\\
60 & 55 & 2031169 & 227445 & 	  49608 &     8489 &  1207517 &  55161& 187254&  12923&  56759&   5317& 390611&  17457&  21534&   2264&   9850&   1071&   3576&   1838&   1410\\
65 & 60 & 1617230 & 193927 & 	  41370 &     7220 &  1058375 &  50788& 173459&  12127&  52661&   5039& 332166&  15743&  19066&   2203&   8630&   1048&   3063&   1531&   1197\\
70 & 65 & 1419335 & 168945 & 	  36062 &     6064 &   927973 &  42934& 151376&  10316&  44148&   4136& 278120&  12937&  16547&   1808&   7447&    822&   2578&   1325&   1022\\
75 & 70 & 1204662 & 135513 & 	  29867 &     4605 &   792486 &  37015& 121558&   8221&  34147&   3190& 227147&  10912&  13190&   1494&   5749&    675&   2127&   1033&    792\\
80 & 75 & 860091 & 88480 & 	  19796 &     2865 &   563957 &  27906&  78969&   5847&  21026&   2238& 155680&   8158&   8642&   1069&   3668&    475&   1370&    645&    512\\
85 & 80 & 471126 & 59667 & 	  12490 &     1978 &   310436 &  15381&  53200&   3836&  14757&   1510&  87185&   4318&   5673&    638&   2445&    290&    842&    401&    307\\
90 & 85 & 164322 & 20454 & 	   4760 &      795 &   107786 &   5417&  18269&   1330&   5094&    523&  27164&   1384&   1682&    212&    754&    102&    267&    138&    119\\
-90 & 90 & 65266291 & 5808556 & 1253280 &   290790 & 13704403 & 753888& 1868061& 149944& 590692&  64510& 5926762& 340811& 286779&  35621& 139046&  17696&  51012&  28319&  21606\\
\hline
\end{tabular}
}
~\\
\medskip{Note: The columns ``rank0'' give the number of UV sources with one SDSS counterpart (within the match radius of 3\as), and ``rank1''
of those with multiple optical matches. UV sources with multiple matches may have composite UV colors, therefore are excluded from the analysis sample; their exclusion
is accounted for by correcting the density of sources by (rank1/(rank0+rank1)). The fraction from columns 16-17 (from which our analysis sample is extracted) is plotted in Fig. \ref{f_exptime2}.
Numbers in columns 18, 19 and 20 include only ``rank0'' sources. }  
\end{table}
}
\end{landscape}

\end{document}